\documentclass[fleqn,usenatbib]{mnras}

\usepackage{newtxtext,newtxmath}

\usepackage[T1]{fontenc}

\DeclareRobustCommand{\VAN}[3]{#2}
\let\VANthebibliography\thebibliography
\def\thebibliography{\DeclareRobustCommand{\VAN}[3]{##3}\VANthebibliography}

\usepackage{graphicx}	
\usepackage{amssymb}
\usepackage{amsmath}
\usepackage{mathtools}
\usepackage{float}
\usepackage{hyperref}

\def \RRI {Raman Research Institute, C. V. Raman Avenue, Sadashivanagar, Bengaluru 560080, India}
\def \kunits {$h~{\rm Mpc}^{-1}$}
\def \pkunits {$({\rm mK})^{2}(h^{-1}\rm Mpc)^{3}$}

\title[EoR drift scan data from MWA]{Extracting the 21~cm  EoR signal using MWA drift scan  data}

\author[A. K. Patwa et al.]{
Akash Kumar Patwa$^{1}$\thanks{E-mail: akpatwa@rri.res.in (AKP)},
Shiv Sethi$^{1}$,
and K. S. Dwarakanath$^{1}$
\\
$^{1}$\RRI
}

\date{Accepted 2021 April   5. Received 2021 March    3; in original form 2020 December 18} 

\pubyear{2020}

\begin{document}
\label{firstpage}
\pagerange{\pageref{firstpage}--\pageref{lastpage}}
\maketitle

\begin{abstract}
The detection of redshifted hyperfine line of neutral hydrogen (HI)  is the most promising 
  probe of the Epoch of Reionization (EoR). We report  an analysis
  of 55~hours of Murchison Widefield Array (MWA) 
  Phase~II drift scan EoR data. The data correspond
  to a central frequency $\nu_0 = 154.24 \, \rm MHz$ ($z\simeq 8.2$ for the
  redshifted HI hyperfine line) and bandwidth $B = 10.24 \, \rm MHz$. As one expects greater system 
  stability in a drift scan, we test the system 
stability by comparing the extracted power spectra from data with noise simulations
and show that the power spectra for the cleanest data 
behave as thermal noise. We  compute the   HI power spectrum  
as a function of time in  one and two dimensions. 
The best upper limit on the one-dimensional  power 
spectrum  are:  $\Delta^2(k) \simeq (1000~\rm mK)^2$ at $k \simeq 0.2${$h~{\rm Mpc}^{-1}$}
and at $k \simeq 1${$h~{\rm Mpc}^{-1}$}. 
The cleanest modes,  which might be the most suited for obtaining the  optimal 
signal-to-noise, correspond to $k \gtrsim 1${$h~{\rm Mpc}^{-1}$}. 
We also study the time-dependence of the foreground-dominated modes
in a drift scan and compare with the expected behaviour. 
\end{abstract}

\begin{keywords}
cosmology: observation; 
early Universe; 
dark ages, reionization, first stars;
techniques: interferometric; 
methods: observational; 
methods: data analysis
\end{keywords}

\section{Introduction}
The probe of the end of cosmic dark age remains an outstanding issue in modern cosmology.  
From theoretical considerations, we expect the  first luminous objects to appear at a  redshift 
$z \simeq 30$. The ultraviolet and other  radiation from these first sources ionized and heated  
the neutral hydrogen (HI)  in their neighbourhood. As the universe evolved,  these ionized regions  
grew and merged, resulting in a fully ionized universe by $z\simeq 6$, as suggested by the 
measurements of Gunn-Peterson troughs of quasars (\citealt{fan06}). Recent Planck  results 
on cosmic microwave background (CMB) temperature and polarization anisotropies fix  the reionization 
epoch    at   $z \simeq  7.8$ (\citealt{planck18}).   The cosmic time  between the  formation of 
the first light sources ($z \simeq 30$, the era of cosmic dawn) and the universe becoming 
fully ionized ($z \simeq 6$) is generally referred to  as the  Epoch of Reionization (EoR). 

Many important astrophysical  processes during this era, e.g.   the formation of first light sources 
and the  evolution of ionized regions around them,  can be  best probed using the hyperfine 
transition of  the neutral hydrogen atom.  Due to the expansion of the universe, this line of 
rest frame frequency $\nu \simeq 1.4 \, \rm GHz$,  redshifts   to frequencies  70--200~MHz 
($z \simeq 6\hbox{--}20$), which can be detected using meter-wave radio telescopes.

Several existing and upcoming radio telescopes aim to detect 
both the sky-averaged and the fluctuating 
component of redshifted HI signal, e.g. radio interferometers---Murchison Widefield Array (MWA 
\citealt{tingay13_mwasystem}, \citealt{bowman13_mwascience}), Low Frequency Array (LOFAR 
\citealt{vanhaarlem13}), Donald C. 
Backer Precision Array for Probing the Epoch of Reionization (PAPER 
\citealt{parsons14}), Hydrogen Epoch of Reionization Array (HERA \citealt{deboer17}), and the Giant Metrewave Radio 
Telescope (GMRT, \citealt{GMRT2013}). In addition there are multiple ongoing experiments to detect the 
global HI signal from this era---e.g. EDGES and  SARAS (\citealt{bowman18}, \citealt{singh18}).

We focus on the fluctuating component of the HI signal in this paper. 
There are considerable difficulties in 
the detection of this signal.   Theoretical studies suggest  that the strength of this  signal  is of the 
order of 10~mK at $150~\rm MHz$ while  the foregrounds are brighter than 
100~K (for detailed review see 
\citealt{furlanetto06}, \citealt{morales10}, \citealt{21cm_21cen}). These  foreground
 contaminants  include the diffuse 
galactic synchrotron  emission and the extragalactic radio sources.  
Current experiments can 
reduce the thermal noise of the system to suitable levels in many hundreds of   hours of integration. The 
foregrounds can potentially be mitigated by using the fact that  the  HI signal and 
its correlations emanate 
from the  three-dimensional structures of mega-parsec scales  at high redshifts. On the the other hand, the  
foreground contamination is dominated by spectrally smooth sources.  This means that even 
if foregrounds can mimic the HI signal on the plane of the sky, the third axis, corresponding to the 
frequency, can be used to distinguish between the two. All ongoing experiments exploit this spectral 
distinction  to isolate the  HI signal from the foreground contamination (e.g. \citealt{parsons09}, 
\citealt{parsons12b}).

Many image  and visibility-based pipelines have been developed to analyze
the interferometric data. These have yielded 
upper limits on the HI signal  for the EoR
(\citealt{GMRT2013, dillon15, paul16, beardsley16, choudhuri16, trottchips2016, patil17, 
paper_reanalysis, 2019ApJ...887..141L, barry19_newlim, 2020MNRAS.493.1662M, 2020MNRAS.493.4711T}). 
The current  best upper limits on the HI power spectrum lie in the range:  
$\simeq (50~\rm mK)^2 \hbox{--} (75~\rm mK)^2$ 
for $k\simeq 0.05\hbox{--}0.2$\kunits in the redshift range
$6.5\hbox{--}10$.  More recently,  \cite{bowman18} reported the  
detection of  an absorption trough of 
strength $\simeq 500~{\rm mK}$ in the  global HI signal in the redshift range $15 < z < 19$.

Some of the key requirements for the detection of the  weak  HI signal are extreme stability of 
 the system, precise calibration, and reliable isolation of foregrounds. Drift 
 scans  constitute a   powerful method   to achieve instrumental stability
 during an  observational run. During such a scan the primary beam and other instrumental 
 parameters remain unchanged while the sky  intensity pattern  changes.
Two interferometers, 
PAPER (PAPER is no longer operational)  and HERA, work mainly in this mode while
 other telescopes (e.g.~MWA) can also acquire data in the drift scan mode. 
 Different variants of drift scans have been proposed in the literature: 
$m$-mode analysis (\citealt{m-mode1, m-mode2}, applied to OVRO-LWA data in
\citealt{eastwood18}), cross-correlation of the HI signal in time (\citealt{paul14}), drift
and shift method (\citealt{trott14}) and fringe-rate  method
(\citealt{fringe-rate}, applied to PAPER data).

 In this paper,  we present the results of the analysis of 55~hours of drift scan  MWA Phase~II data.  
 The data was taken over 10 nights with repeated scans of duration 5.5~hours  over
 the same region of the sky.  The feasibility of drift scan using MWA has 
 been studied theoretically (\citealt{trott14, paul14, patwa19}).  In this paper,
 we use the formalism developed by \cite{patwa19}. We work in  delay space,  
 which isolates foregrounds from the EoR window {and} is therefore 
the natural domain of power 
 spectrum measurement (e.g. \citealt{datta10}, \citealt{parsons12b}).

We develop a pipeline for measuring the 21~cm HI power spectrum for drift scans in 
the delay space.  
In section~\ref{sec:data}, we describe   the  MWA phase  II EoR data and its pre-processing, 
e.g. flagging, time-averaging, and processing using 
Common Astronomy Software Application\footnote{CASA: \url{https://casa.nrao.edu/}}
(CASA; \citealt{2007ASPC..376..127M}). In section~\ref{sec:data_meth}, we
briefly review  different methods of extracting the HI signal from drift scan data  and 
provide justification for  the method adopted in this paper. In section~\ref{sec:dataana},
we discuss the  transformation of  the gridded data  to the delay space and
outline  techniques  needed  for 
the estimation of the HI power spectrum from the data. In section~\ref{sec:res_star}, 
the main results are presented. 
In this section, we first characterize  the noise properties 
of the data and study the  stability of the system over the duration of the scan  by comparing 
the data power spectrum with  noise simulations. In section~\ref{sec:h1_powspec},   
we apply our methods to   extract the 
HI power spectra  in  different representations of the HI signal  in Fourier space. 
Finally, the behaviour  of the foreground-dominated modes of the data is discussed 
in a drift scan. We summarize our results and conclude with possible future directions of our work 
in Section~\ref{sec:concl}. 

Throughout this paper we use the spatially-flat  $\Lambda$CDM model for our work with values  
$ h = 0.67$, $\Omega_\Lambda = 0.6911$, $\Omega_m = 0.3089$, $\Omega_k = 0.0$  (\cite{planck18}).

\section{data}\label{sec:data}
\begin{figure}
		\includegraphics[width=\linewidth]{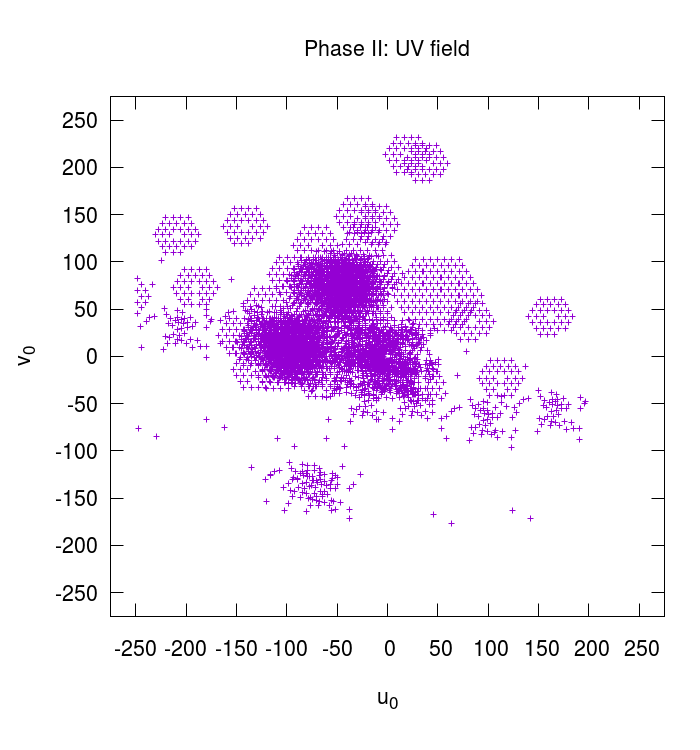}
	\caption{MWA  Phase~II  baseline distribution for a zenith drift  scan  
	(in the units of central wavelengths $\lambda_0 = 1.95~\rm m$)}
        \label{fig:ph2_uvfield}
\end{figure}

Murchison Widefield Array (MWA) is a radio interferometer 
located in  Western Australia at a 
latitude of  $-26.7^\circ$ and longitude of $116.7^\circ$.
It  has 128 tiles (a tile would  be referred to as an antenna in the paper) and 
each tile consists of 16 crossed dipoles 
placed on a square mesh of length  $\sim4~{\rm m}$ in a 4x4 arrangement.
It operates in the  frequency range of 80--300~MHz with an
observational  bandwidth of 30~MHz  
and   a frequency resolution (channel width) of 40~kHz.
The  visibility data is recorded at  an  interval of 2~minutes with the temporal resolution of 0.5~s.
In each 2-minute snapshot, the duration of observation is  112~s.
In the Phase~II design  of  MWA, 64 antennas are placed in a compact Hex configuration 
to increase the number of short and redundant baselines primarily for  EoR studies. 
The instantaneous baseline distribution of MWA Phase~II for a zenith scan  is shown in 
Figure~\ref{fig:ph2_uvfield} 
(e.g. See \citealt{tingay13_mwasystem, mwaphaseII_design} for a  detailed description of the MWA design).

 \subsection{Metadata}\label{sec:metadata}
\begin{figure}
	\centering
	\includegraphics[width=\linewidth]{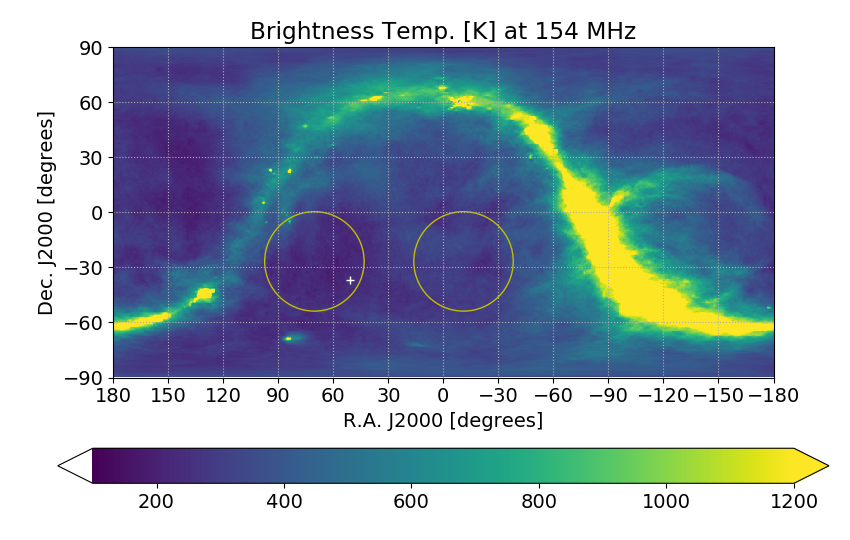}
	\caption{This figure displays the  Haslam map scaled to 154~MHz 
	  assuming the brightness temperature spectral index $\alpha = -2.52$ (\citealt{rogers08}).   
	  The big circles denote   the main lobe of the MWA primary beam at $\nu = 154 \, \rm MHz$.  
	  The scan starts roughly at the location of the circle on the right and lasts until 
	  the field of view corresponds to  the circle on the left. The white `+' sign shows 
	  the  position  of Fornax~A.}
	\label{fig:sky_covered}
\end{figure}

 In Figure~\ref{fig:sky_covered} we display the region of
 the sky covered by  the drift  scans  with the MWA. 
 {The drift scan  data we use in this paper was acquired
 in 2016 \citep{2016mwa..prop..B08P} and at the time of the analysis was publicly available
 in the MWA Data Archive\footnote{MWA Data Archive: \url{https://asvo.mwatelescope.org/}}
 with project ID G0031.} 
 Each scan  lasted  5 hrs 24 minutes and was repeated 
 across the same region of the sky for 10 consecutive nights from 2016-Oct-03 to 2016-Oct-12. 
The scanned region covers a portion of the sky   
from the position of the  EoR0  
($0{\rm h}, -27^\circ$) to the EoR1   
($4{\rm h}, -27^\circ$) 
fields of the MWA.
 Each night the 
 scan was carried out  between 14:39~UTC and 19:28~UTC. On the sixth night, nearly two hours of data was missing  during the initial period and the scan lasted only 3 hrs 10 minutes.

 The reported  sky temperatures in the regions covered by the  scan  vary in
 the range $210$--$267$~K.
The scan also passes over Fornax~A which is an extended radio source 
with a core and two radio lobes. Its angular extent spans the region, 
RA $3{\rm h}25{\rm m}$ to $3{\rm h}20{\rm m}30{\rm s}$ and Dec $-37^\circ 30'$ to $-36^\circ 54'$. 
The total flux density of this source at 154~MHz is 750~Jy (\citealt{mckinley15}).

\subsection{Flagging, calibration, and averaging}\label{sec:calib}
COTTER is a pre-processing pipeline which flags RFI, non-working antennas and 
frequency channels (\citealt{offringa15}). 
It also does cable correction, can average data  and produce CASA readable 
\textit{Measurement Sets} (called MS tables/files). 
We apply COTTER on all 2~minute snapshots individually and average them to 10~s time resolution, 
with the frequency resolution kept intact (40~kHz).  
{The MWA publicly-available archive provides COTTER to enable the reduction 
in the data volume. COTTER was applied on-site before the data was downloaded.} 
In each 2-minute data file, as noted above, the  observation span is  112s. 
Thus,  after time-averaging there are 11 data chunks in each 2-minute data file.
We then apply  the bandpass and the flux density calibration with the  strong and unresolved calibrator 
source Pictor~A.\footnote{
Based on \cite{jacobs13},  Pictor A corresponds to:
(RA, Dec) = ($5{\rm h}20{\rm m}22{\rm s}, -45.8^\circ$), 
flux density $S_\nu = (381.88 \pm 5.36)$~Jy at 150MHz, and 
flux spectral index $\alpha = -0.76 \pm 0.01$.}

The calibration solution tables also help in identifying and flagging 
unresponsive or irregularly behaving antennas and baselines. 
Next we flag the channels situated at 
either ends of the coarse bands of MWA bandpass.\footnote{MWA's  frequency band consists of 24 coarse bands, giving  a
total observing bandwidth of 30.72 MHz. 
Each coarse band has 32 channels. MWA by design has missing frequency channels in each coarse band. 
In all data files, we flag 4 channels at both ends and 1 channel at the
center of each coarse band. 
That is if we number channels from 0 to 31, channel number 0, 1, 2, 3, 16, 28, 29, 30, 31 are flagged.}

These processes yield calibrated  time-ordered  visibility data-sets of 10~nights. 
We then  performed   averaging over data from different  nights. 
Since the  drift scans cover  the same region of the sky, we carry out LST stacking 
(this is the usual procedure to  add data for  all transit radio telescopes e.g. \cite{chime}) 
which constitutes  aligning and averaging  data snapshots with the  same 
tracking centers, baseline bins (see below for more details), and frequency channels observed 
on different nights. 
After LST stacking, we obtain the equivalent of one night
 of drift scan data over the same region of 
the sky. These data  are  suitable  for the EoR power spectrum analysis.

\section{Data analysis methodology} \label{sec:data_meth}
Many operational  radio interferometers rely on drift
scans to extract the power spectrum of the intensity of  redshifted HI line from the  EoR and low
redshift data (\citealt{chime, paper_reanalysis, fringe-rate, deboer17}). 
CHIME adopts  m-mode decomposition 
of the time-ordered visibility data, which relies upon Fourier transforming the data stream.
PAPER and HERA use weighted averages of the visibility data. In \cite{patwa19} (hereafter PS19),
we proposed many  different approaches to analysing the drift
scan data and showed similarities and differences between existing methods.  In this paper, 
we adopt a method based on cross-correlating the time-ordered visibilities (for details see PS19).

The main aim of all the analysis pipelines  is  to construct  an unbiased and optimal estimator to extract 
the  HI power spectrum from the visibility data in frequency or delay space. 
Unlike the tracking data, the intensity pattern in a drift scan
changes with respect to the beam, and the analysis of these time-dependent visibilities arising from
a changing intensity pattern presents new challenges.

All the  methods of extracting the HI power spectrum directly from the visibility
data are based on the correlation properties of the measured visibility
$V_\nu(\textbf{u}_\nu,w_\nu,t)$ in different domains. These properties have
been well studied for the tracking data for frequency and baseline domains
and can readily be extended to the drift scan data (e.g. PS19).  Here our
focus is the correlation of measured  visibilities in the time domain. PS19 derived the
de-correlation profile for the primary beams of many operational  and
future interferometers. Even though this profile is a  complicated  function of the 
baseline length, the de-correlation time varies between a  few minutes to 30 minutes 
for most interferometers.
As the visibility data generally has   higher time resolution as compared to the de-correlation time, 
e.g. we use
MWA data with 10~second resolution  in this paper, multiple methods
can be used to analyse the data. In the analysis of PAPER data, a time filter
is used to average over visibilities which contribute coherently to the HI signal 
(e.g. \citealt{paper_reanalysis} uses 43-second time filters). As already noted above, 
CHIME data analysis is based on Fourier transforming the time-ordered
visibility data (for details see PS19). 

In our work, we adopt the method based on cross-correlation of visibilities
in time (PS19). This method is applicable to both frequency and delay space
data. In this work, we transform visibility to delay space to isolate foregrounds. 
The HI power spectrum  is extracted using the estimator (Eq.~(45) of PS19)
for $t\ne t'$ to avoid noise bias:
\begin{align}
  {\cal C}_\tau&(\textbf{u}_{0},w_0) \nonumber \\
  &= \frac{\sum_{t', t} \exp(-i2\pi u_0 \cos{\phi} \Delta H) V_
  \tau(\textbf{u}_{0},w_0,t) V_\tau(\textbf{u}_{0},w_0,t')^* g(t'-t)} {\sum_{t', t} g^2(t'-t)}
  \label{eq:h1estim}
\end{align}
Here  $g(t'-t)$ is the function that captures the de-correlation of visibilities
as a function of the  time difference $t'-t$ (Figure~1 of PS19). $g(t'-t)$
is nearly unity for $t' - t < 20$~minutes for the shortest baselines we consider in this paper, 
$\sqrt{u^2+v^2} \simeq 20$ \footnote{We note that the shortest available baselines for MWA corresponds 
to $\simeq 4\lambda_0$. However, we exclude them as the signal for these baselines is heavily 
contaminated owing to instrumental systematics.} 
and the de-correlation  time scale falls to around 5~minutes for
the longest baselines, $\sqrt{u^2+v^2} \simeq 300$.  $\Delta H$ is the difference of the hour 
angle between times $t$ and $t'$. $\tau$ is the delay
space parameter and $\textbf{u}_{0} \equiv (u_0,v_0)$ and $w_0$ are the baselines
and the $w$-term at the center of the bandpass.
As shown in PS19, this estimator is both unbiased and optimal for the extraction of the HI 
signal\footnote{Eq.~(\ref{eq:h1estim}) can be understood
  more easily by assuming  $g(t'-t)$ to be unity for $t'-t \le t_0$, where $t_0$ is some fixed 
  time that depends on the baseline, and zero for $t'-t>t_0$. All the cross-correlation for  
  $t'-t \le t_0$ can be  used for computing the HI power spectrum
  (or equivalently visibilities can be averaged over this time interval using
  a filter e.g. \cite{paper_reanalysis}). This process will yield an unbiased and optimal estimator.  
  If the time over which the visibilities
are cross-correlated is shorter than $t_0$, then the estimator is still unbiased but it is not optimal. 
If the time is chosen to be longer than $t_0$, then the HI signal gets uncorrelated and consequently, 
the estimator is neither unbiased nor optimal.}. 
The measured quantity (Eq.~(\ref{eq:h1estim})) is converted
to the variables of the HI signal using relations given in  Appendix~\ref{sec:appen}.

\section{Data Analysis} \label{sec:dataana}
  The visibilities 
in delay space can be derived from the visibilities in the frequency space by performing 
a discrete Fourier transform:
\begin{equation}
V_{\tau}(\textbf{u}_{0}, w_{0},t) = \Delta \sum_{\nu_0 - \frac{B}{2}}^{\nu_0 + \frac{B}{2}}  V_{\nu}
(\textbf{u}_{\nu}, w_{\nu},t) B_\nu e^{2 \pi i \tau \nu}.
\label{BN_win}
\end{equation}
For our analysis, the central frequency, $\nu_0 = 154.24 \, \rm MHz$ and $B= 10.24 \, \rm MHz$, 
which yields  256 channels of channel width $\Delta = 40 \, \rm KHz$.  $B_\nu$ corresponds to 
Blackman-Nuttall window (\citealt{nuttall81}), which helps reduce power leakage between delay bins.  
 Our analysis is entirely based on analysing the  
visibilities and their correlations and at no stage do we 
transform to  the image domain. 
We also do not subtract  foreground sources based on any model of foregrounds but rely on 
the separation of foregrounds 
from the EoR signal in delay space.

\subsection{Gridding of UV field and power spectrum estimation}\label{sec:uvfield_griding}
In Figure~\ref{fig:ph2_uvfield}, we display the baseline distribution of
MWA Phase~II for our observational setting (zenith scan).  
In a drift scan, the  UV distribution and the $w$-term 
are  left unchanged. For zenith scans, the $w$-term is generally
small,  $|w_0| \leq 3 \lambda_0$ for MWA,
and its impact on the interpretation of
data can be neglected (for details e.g. see PS19). We select a square UV field with $u_0$ and $v_0$ 
in the range:  $-250 \lambda_0$ to $+250 \lambda_0$.
This allows us to include   all the UV  regions in which
the density of baselines is large for MWA Phase~II (Figure~\ref{fig:ph2_uvfield}).

For analysing the data, we grid the UV field, with the size of a pixel determined by the expected 
behaviour of  the HI signal. It can readily be shown
that, for MWA,  the HI signal de-correlates for 
baselines that  differ by  more than a
few wavelengths (e.g. \citealt{morales04}, \citealt{paul16} 
and references therein).  We choose a square  pixel of  $0.5 \lambda_0$ so that the equal-time visibilities 
are  coherent inside a grid; the baselines are assigned the same value inside a pixel. We also assume 
inter-pixel visibilities 
to be uncorrelated.  Each UV pixel also has non-zero width in delay space  ($\simeq 1/B$)   owing 
to finite bandwidth. For every UV grid, the HI signal is coherent within  this width 
(e.g. PS19, \cite{paul14,fringe-rate} and references therein).
We refer to this three-dimensional grid, labelled by
${\bf u}_0, \tau$, as `voxel' in the rest of the paper.

We  populate the  gridded UV field (for a given $\tau$)  with  visibility data 
(each data point corresponds to an  integration time  $t_{\rm int} = 10$~s).  
For the combined data of 10 nights, the number of voxels  and the maximum occupancy of a voxel 
for a fixed $\tau$
and time  are 5427 and 55, 
respectively (Figure~\ref{fig:ph2_uvfield}).

The correlation function (Eq.~(\ref{eq:h1estim})) (and  the power spectrum using Eq.~(\ref{eq:defpwsapprox2}))  can be computed  
from the gridded data for a given voxel.  
This procedure can be extended  to further compute mean power spectra, by carrying out
a weighted average (the weight can be theoretically computed
from the occupancy  of a given voxel): (a) over all the voxels for a fixed $\tau$,  (b) over 
a set of voxels for 
a given $k_\parallel$ and $|{\bf k_\perp}|$,  or (c)  a set of voxels for a fixed 
$|{\bf k}| =\sqrt{{\bf k_\perp}^2+k_\parallel^2}$.
All these quantities can be estimated as a function of integration (or drift) 
time\footnote{In this paper we use terms integration time and drift time interchangeably and plot the 
power spectra and their RMS as a function of drift time. These concepts do not necessarily mean 
the same thing and therefore further clarification is needed. The occupation of a grid increases 
with the time of the drift scan. For a scan of duration $t'$ all 
cross-correlations of time difference $\Delta t < t'$ are included in our analysis. 
An increase in the number of realizations (cross-correlation plus incoherent averaging over 
different voxels)  causes,  for noise-dominated data,   a decrease of 
the mean power spectrum and its RMS as a function of time  which is similar to the outcome of  
integrating longer  in  
a tracking observation. One case in which drift time and integration time
differ is when there is missing data, e.g. on the sixth night, two hours
of  data is missing. In all the figures, the x-axis denotes the drift time.}. These averages are in 
units of $(\rm JyHz)^2$; they are converted to \pkunits
 using Eq.~(\ref{eq:defpwsapprox2}).   The estimated  power spectrum is a complex number.  
Throughout this paper, while displaying the power spectrum, we plot
the absolute value of this complex number.

\section{Results} \label{sec:res_star}
Given that the expected HI signal can 
only be detected after  hundreds of hours of integration, 
it is imperative that the noise of the instrument is characterized precisely---this allows us to gauge  
the  stability of the system, e.g. primary beam and bandpass, and the extent of  foreground contamination.
In this paper, we analyze  55 hours  of drift
scan data in delay space  and expect the data for a fraction of delay space parameters    to be dominated by noise. We attempt to
verify this hypothesis in  the next sub-section. 
 
\subsection{Noise characteristics}\label{sec:noise_charac}
There are multiple ways to compare the data with noise simulations. The most straightforward method would 
be to compare the HI power spectrum extracted from data against  simulated visibilities. We do not 
adopt this method for the following reason.
If there are $N$ visibility measurements in a pixel, there are  a total
of $N(N-1)/2$ cross-correlations. However, the HI power spectrum is based
on only a fraction of these cross-correlations as the HI signal de-correlates
for $t'-t$ exceeding  30~minutes for even the shortest baselines we study here. For noise 
characterization, we give equal
weight to all cross-correlations, which is equivalent to $g(t'-t) = 1$ in
Eq.~(\ref{eq:h1estim}). This also allows us to use the  pipeline developed for
the extraction of HI signal with minor modifications on  the data and simulated visibilities.

For  comparison with data, we simulate visibility data using Gaussian random noise. In this case, 
each visibility cross-correlation has zero mean (because all visibilities
are uncorrelated), and an  RMS $\sigma$ given by (\citealt{ChristiansenHoegbom}):
\begin{align}
  \sigma = \frac{1}{\eta_s} \frac{2k_{B}T_{\rm sys}}{A_{\rm eff}\sqrt{\Delta \nu t_{\rm int}}}.
  \label{eq:noissimu}
\end{align}
For the  simulation, the following  MWA system parameters  are used:   $A_{\rm eff} = 21.4{~\rm m}^2$, $\Delta \nu = 40$ kHz, 
$t_{\rm int} = 10$~s. In addition, we assume  $\eta_s = 1$, and  $T_{\rm sys} = 400$~K. This yields an  RMS 
for a single cross-correlation,  $\sigma = 81.8$~Jy. We note that the comparison of 
noise simulations with data allows
us to determine $T_{\rm sys}/\eta_s$\footnote{Noise simulations
allow us to establish the extent to which the data behaves like thermal 
noise. In the ideal case of equally filled grids we can 
get analytic estimates of the projected noise. Let us assume the total number of visibilities is $N$ (each visibility corresponds to an integration time of 10~s) distributed in $M$ grids, or the occupancy of each grid is $K=N/M$. Let us further assume the signal adds coherently in each grid and incoherently across grids. Neglecting self-correlation, the number of cross-correlations in each grid
are $R\simeq K^2/2$, which gives the expected RMS for  each grid  to be 
$\sigma_{\rm pix} = \sigma/\sqrt{R}$. If these cross-correlations are further averaged incoherently across $M$ grids, the final expected RMS is $\sigma_{\rm fin} = \sigma_{\rm pix}/\sqrt{M}$. We do not reach this noise level for a multitude of theoretical and experimental reasons. First, each cross-correlation inside a grid does not receive the same weight for the expected theoretical  HI signal, as already discussed above. Second, the occupancy of each grid is determined by the baseline distribution of the interferometer (Figure~\ref{fig:ph2_uvfield}) and it is not uniform. Third, we expect foreground contamination  which is expected to increase the RMS above the Gaussian noise.}. We transform these
simulated visibilities to delay domain  for the baseline distribution of MWA Phase~II.

In a given voxel,  the number of cross-correlations increase as the square of the drift time,  $t^2$. If all the cross-correlation are assigned equal weight,  the RMS of the power spectrum computed  from all cross-correlations within a  voxel is expected to scale
as $1/t$. This motivates us to define the following function for  comparing the simulated noise with the data:
\begin{align}\label{eq:f(t)}
	f(t) = a \left(\frac{120 \, {\rm sec}}{t} \right)
\end{align}
where $a$ and $t$ are in units of \pkunits
and seconds, respectively. 
\begin{table}
\begin{center}
	\centering
	\begin{tabular}{ |c|c|c|c| } 
		\hline
		$a~[10^{10}\text{\pkunits}]$  & XX & YY \\
		\hline
		Night 1 &  5.9 & 7.4 \\ 
		Night 2 & 5.3 & 6.9 \\
		Night 3 & 7.3 & 9 \\
		Night 4 & 5.6 & 7.4 \\
		Night 5 & 6.4 & 8.3 \\
		Night 6 & 4.2 & 5.4 \\
		Night 7 & 6.3 & 8 \\
		Night 8 & 5.8 & 7.3 \\
		Night 9 & 6.2 & 8.1 \\
		Night 10 & 5.8 & 7.5 \\
		\hline
		\hline
		Mean & 5.9 & 7.5 \\
		\hline
	\end{tabular} 
\end{center}
\caption{
The table displays the  normalization defined 
  in Eq.~(\ref{eq:f(t)}) for the 10-night data for both XX and YY polarizations.
  }\label{tab:tab2}
\end{table}
In Table~(\ref{tab:tab2}) we display the results of 10 nights of data.  
The coefficient $a$ in Eq.~(\ref{eq:f(t)}) is computed using the first half an hour data on every night for the delay parameters with the lowest RMS. Night~6 data gives a smaller value because there is no
data on that night for the first two hours and the data flow starts from  a slightly cooler 
region of the sky. From simulated visibilities we obtain
$a \simeq 11.7 \times 10^{10}$\pkunits,
which is higher than the data for $T_{\rm sys}/\eta_s = 400 \, \rm K$. 
A comparison with data allows us to infer $250 \, {\rm K} <  T_{\rm sys}/\eta_s < 330 \, \rm {\rm K}$. 
For $\eta_s \simeq 1$, the estimated
system temperature is in good agreement with the reported  range of system temperatures 
in the scanned region of the sky. 

When the data from all the nights is combined, we obtain  $a = 8.4 \times 10^9$\pkunits for XX 
and $a = 10.4 \times 10^9$\pkunits for YY. 
A comparison with the values in Table~(\ref{tab:tab2}) shows that
the improvement for the combined 10~nights of data  is   a factor of  
 7 (XX polarization) and 7.2 (YY polarization) while the decrement under the ideal
conditions would be closer to a  factor of 10. 

\begin{figure*}
		\includegraphics[width=0.32\textwidth]{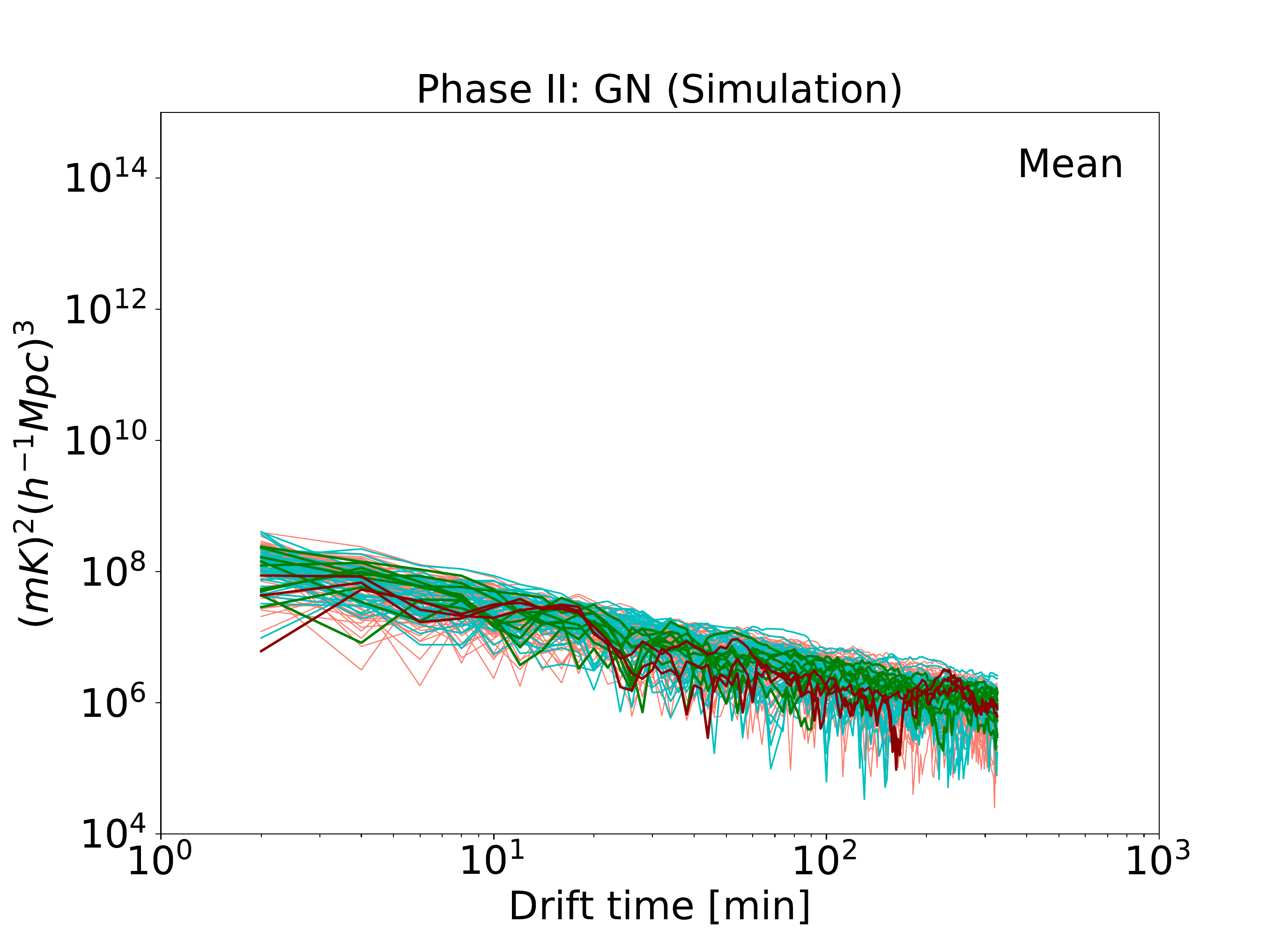}
		\includegraphics[width=0.32\textwidth]{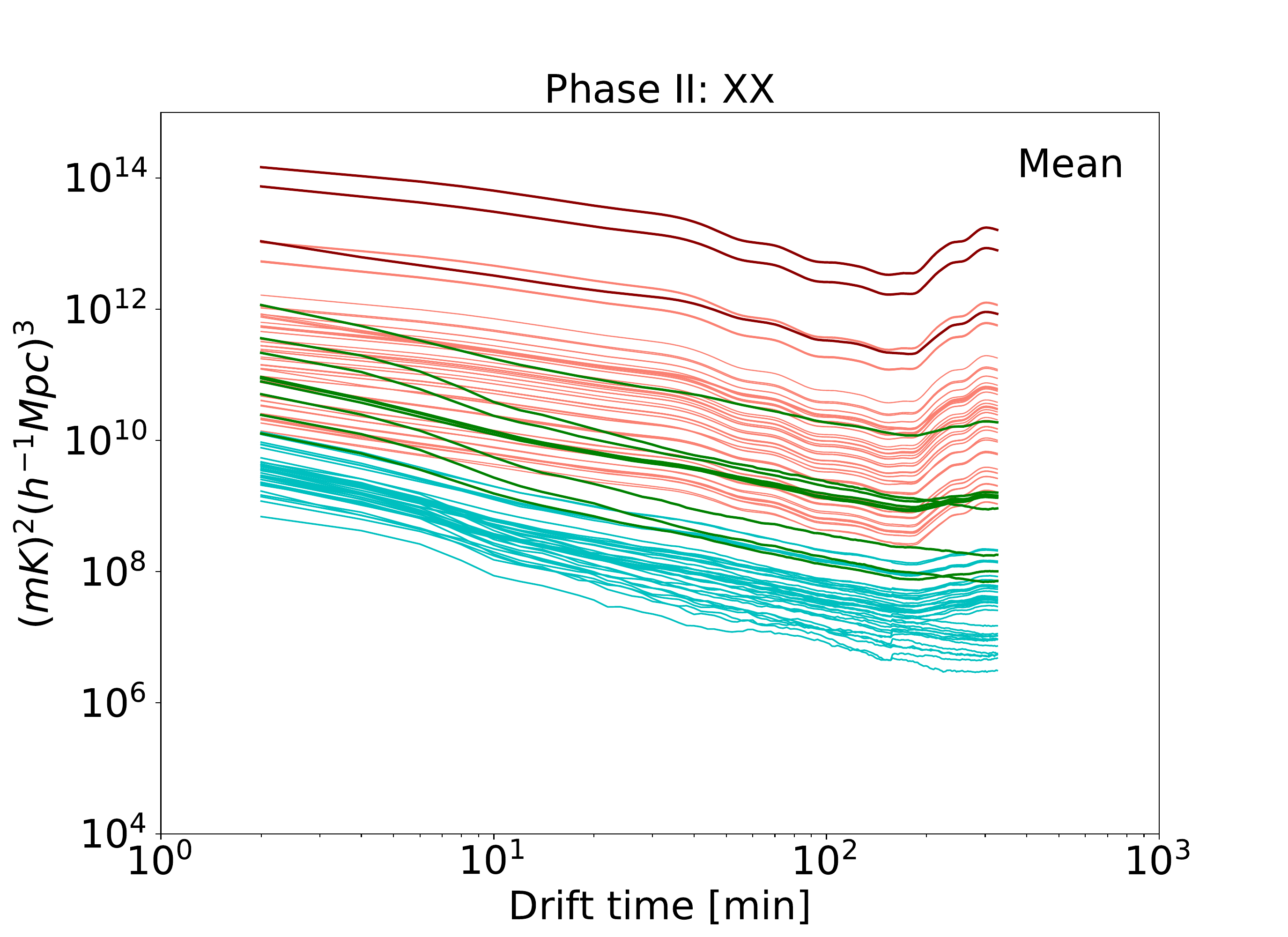}
		\includegraphics[width=0.32\textwidth]{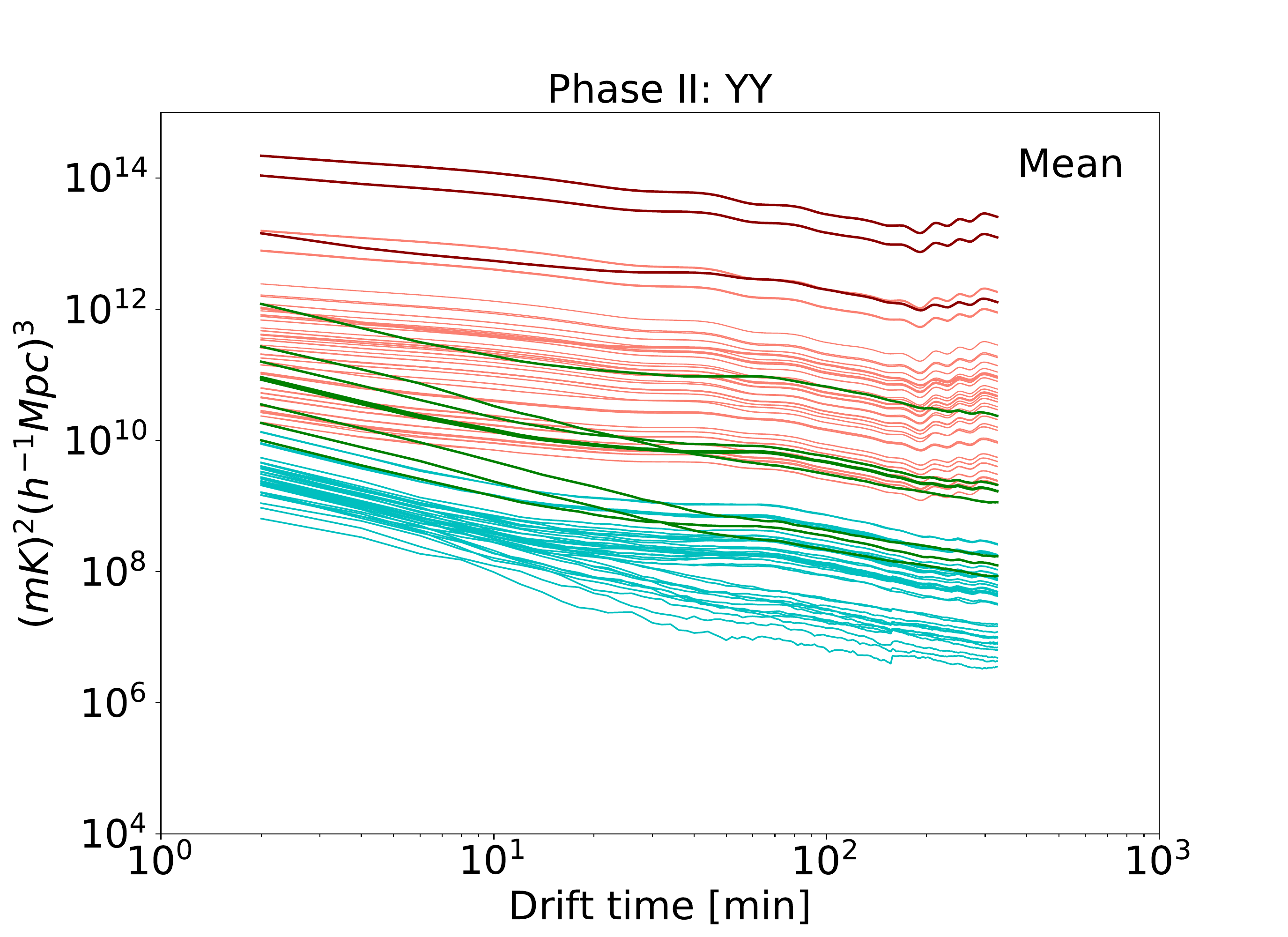}\\
		\includegraphics[width=0.32\textwidth]{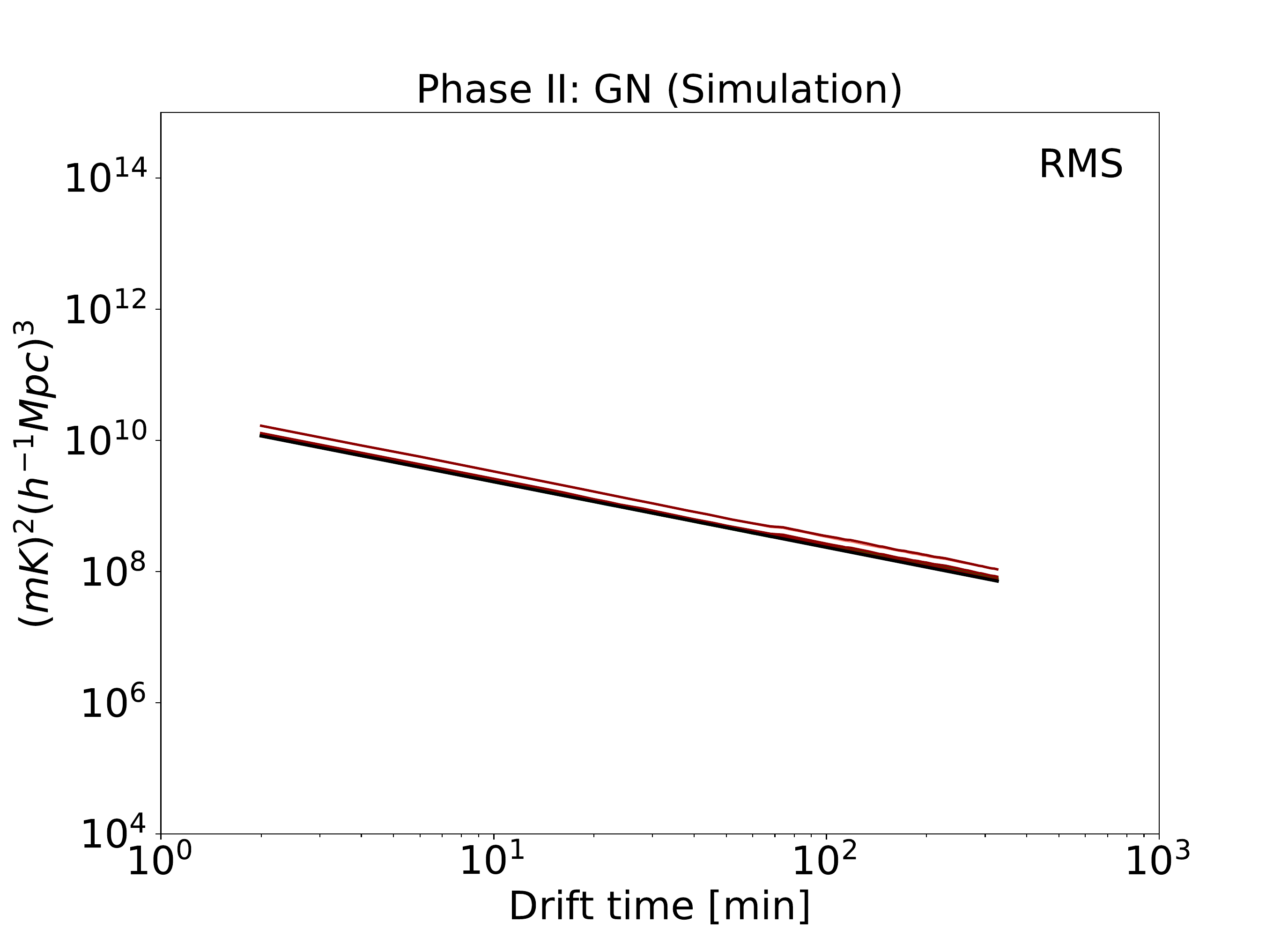}
		\includegraphics[width=0.32\textwidth]{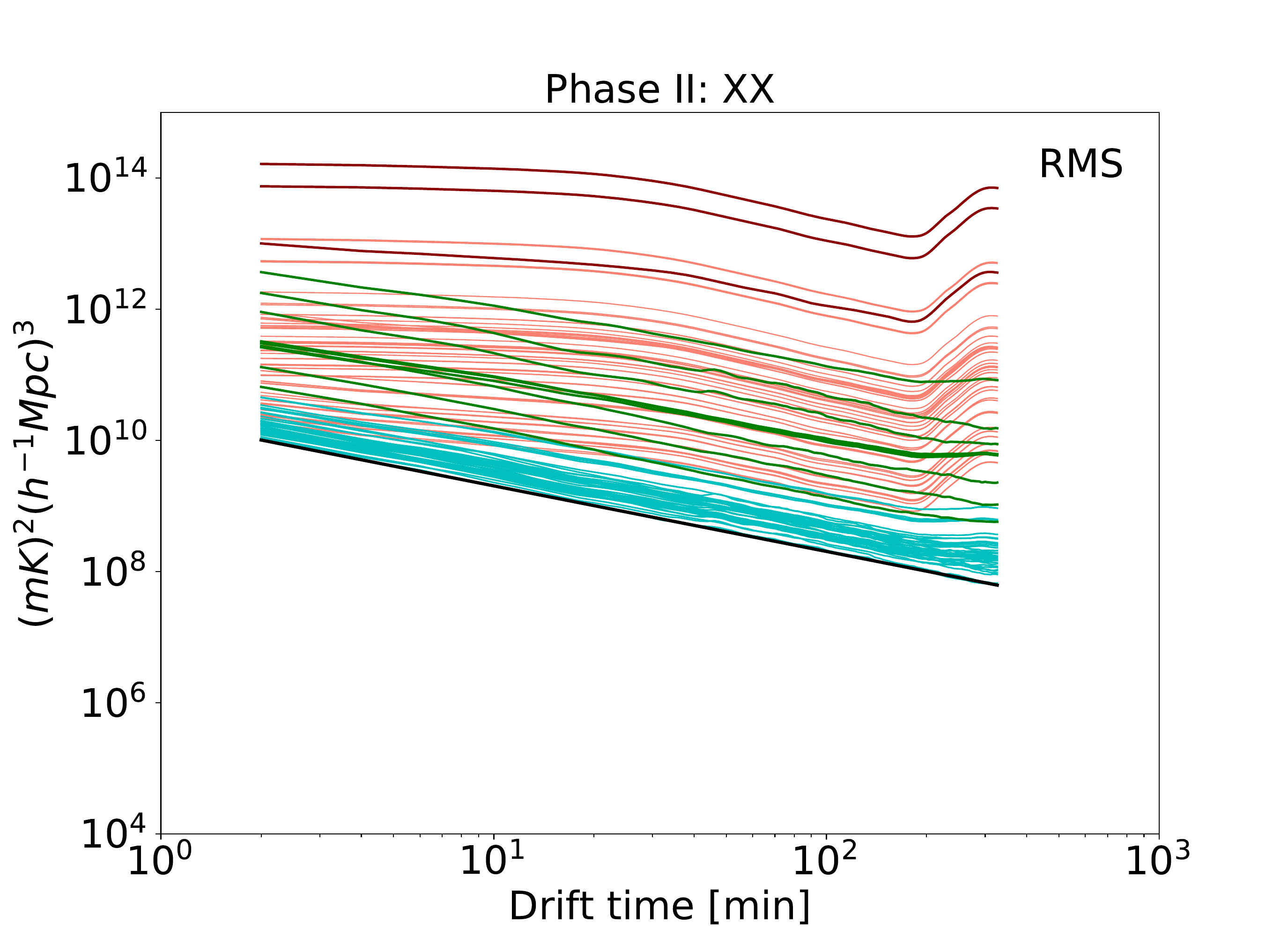}
		\includegraphics[width=0.32\textwidth]{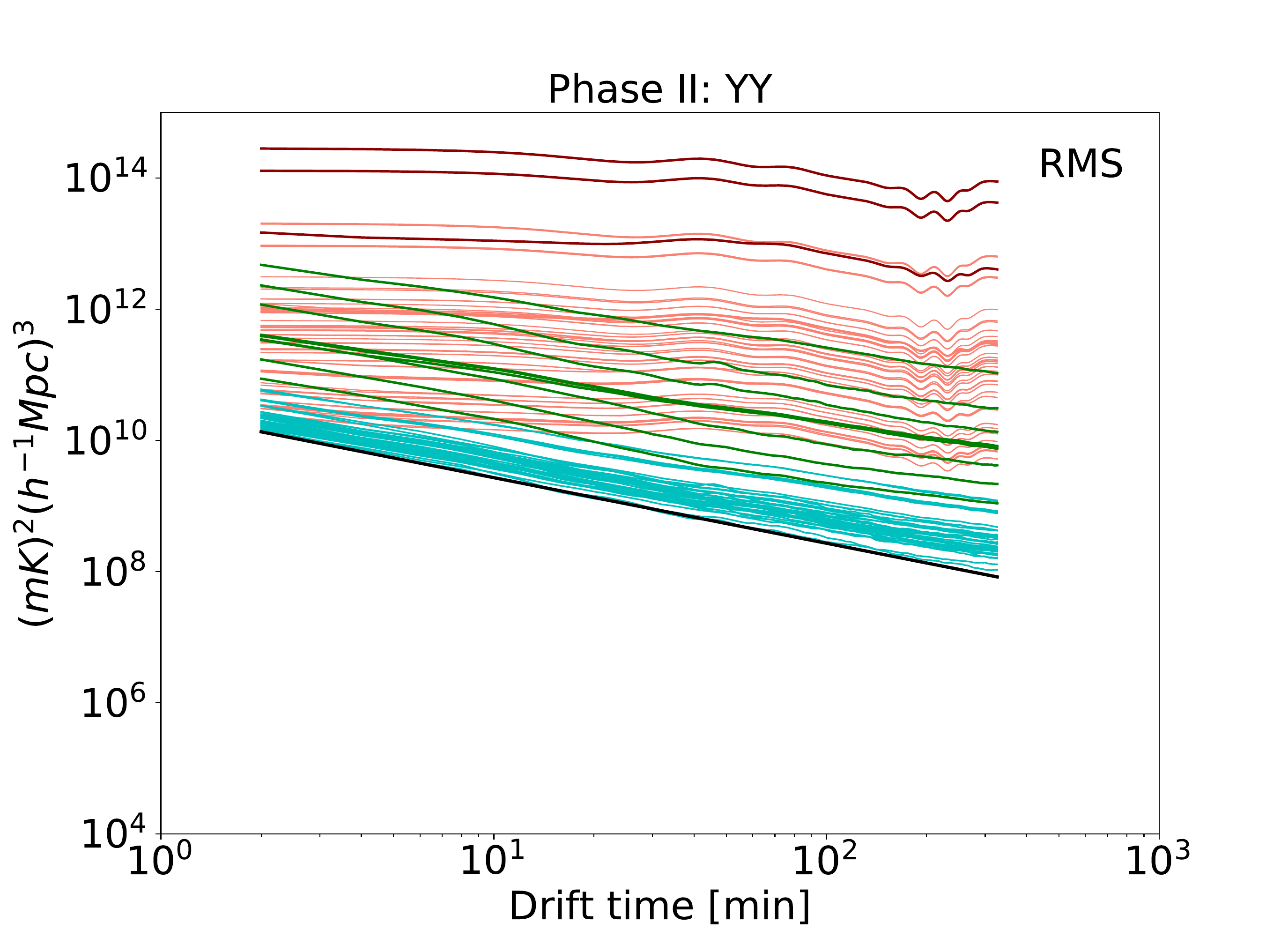}
	\caption{The  mean  power spectra along with the  RMS of power spectra for  one  voxel are  displayed as a  function of drift time (in minutes). 
	The first  column  shows the outcome of  Gaussian noise (GN) simulations for the baseline
distribution of MWA Phase~II. The second and third columns give results for  XX and YY 
polarization, respectively. The first row shows the mean power spectrum while  the second row displays
the RMS of the  power spectra. The  horizontal axis in all the  plots corresponds to  the drift  time. 
There are 129 curves in every panel, each corresponding to a different value of the delay space parameter  
$\tau$ (including $\tau=0$). 
The dark and the light red  coloured curves represent the wedge and 
horizontal bands (including two modes on either side of the band), respectively.
 The rest of the 
 modes in the  EoR window are  divided into  two colours:   intermediate values of $\tau$  
in green   and the clean modes corresponding to  $\tau \ge 22$ (in units of $B^{-1}$) in light blue. 
Thick  black line in the  RMS plots correspond to the function  given in Eq.~(\ref{eq:f(t)}).}
\label{fig:pws_rms_plots}
\end{figure*}

Figure~\ref{fig:pws_rms_plots} shows both  the  simulated noise 
and the data as a function of drift time.  The figures display  the mean power spectra and the RMS of 
mean power spectra for Gaussian Noise (GN) simulations and  the combined data sets of 10 nights. 
Different  lines (129 lines) in
the figures correspond to the values of   the delay parameter $\tau$ (including $\tau = 0$). 
The data is also compared to the 
expected analytic function  (Eq.~(\ref{eq:f(t)})).

Figure~\ref{fig:pws_rms_plots} has been demarcated into several regions to
separate the delay parameters suitable for the  EoR detection from the foreground
dominated modes. This figure should be viewed in conjunction with  the more usual wedge plot
(Figure~\ref{fig:2d_120min_ph2}). The dark red lines correspond to the delay
parameters that are most contaminated by foregrounds ($\tau \la 10/B$). The light red lines
capture the impact of missing channels in MWA. We find the delay parameters
$\tau < 22/B$ to be partially contaminated by foregrounds  and this region is delineated by green
lines. The cleanest regions correspond to light  blue lines which correspond
to delay parameters which are separated from  horizontal bands
(caused  by missing channels) by more than $1/B$  and $\tau > 22/B$. These regions  will be referred
to as `clean modes' in the rest of the paper.  These  modes constitute
25--30\% of the delay parameters. The figure shows that the
signal (the mean and the RMS) could vary by over an order of magnitude even within 
the clean modes, which partly captures the variation of the mean even for
noise dominated modes (the upper left panel of the figure). However, the noise RMS 
clusters around a line (the lower left panel of the figure) which suggests that some of 
the  clean modes could be  affected  by  residual foreground contamination
and other systematic errors. However, Figure~\ref{fig:pws_rms_plots} also shows that 
the  RMS  in the noise  simulation and 
the RMS  in the clean modes from data  (both XX and YY polarizations) decrease as
$1/t$ (Eq.~(\ref{eq:f(t)})). This provides clear evidence that the
clean modes are noise-dominated. 
For  the clean modes, 
there is 
reduction of noise by nearly two orders of magnitude over  4~hrs, as anticipated by the analytic 
fit (Eq.~(\ref{eq:f(t)})). 

The data power spectra begin to  deviate from noise simulations after nearly four hours of drift scan, 
which is related   to   a corresponding increase   in the mean power and the  RMS in 
foreground-dominated modes towards the end of the scan.  This increase can be  understood from 
Figure~\ref{fig:sky_covered}. It is caused when   the  bright   radio source Fornax A enters
the main lobe of the primary beam and  sidelobe
pick up  by the emission from the  galactic plane. 

To test the hypothesis that this enhancement  in the power towards the end of
the scan is caused by extended sources such as Fornax and the galactic plane, we  re-analyse data 
by excluding  smaller  baselines ($<100\lambda_0$).   We find that this procedure removes  the  bump. 
This shows  that  extended sources, even with complex structures, can be removed  if small baselines 
are ignored in power spectra computation. However,  since the HI  signal is expected to be stronger 
on smaller baselines, this procedure is only used for testing and not for
the final analysis of data.

\begin{figure*}
		\includegraphics[width=0.48\textwidth]{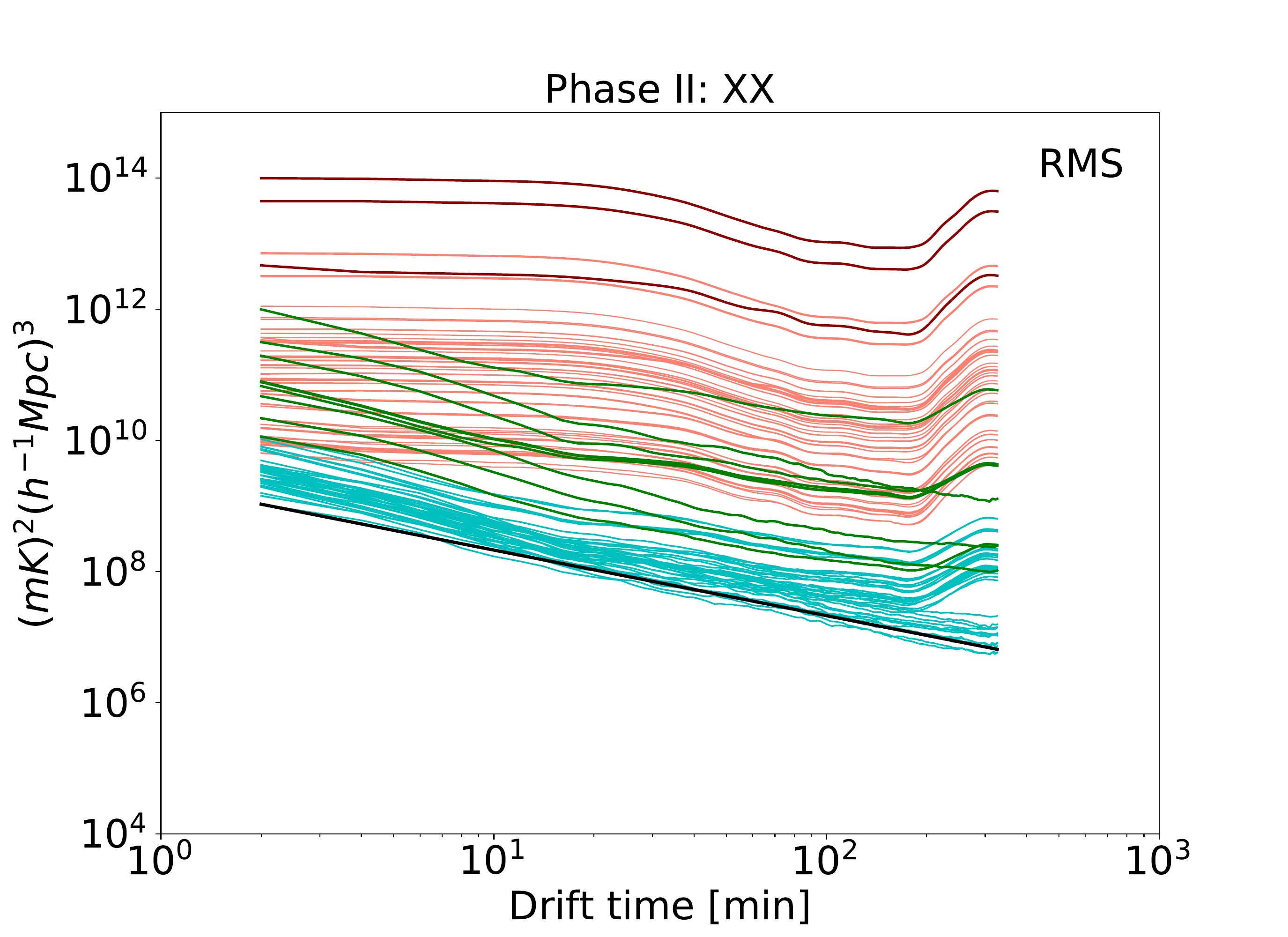}
		\includegraphics[width=0.48\textwidth]{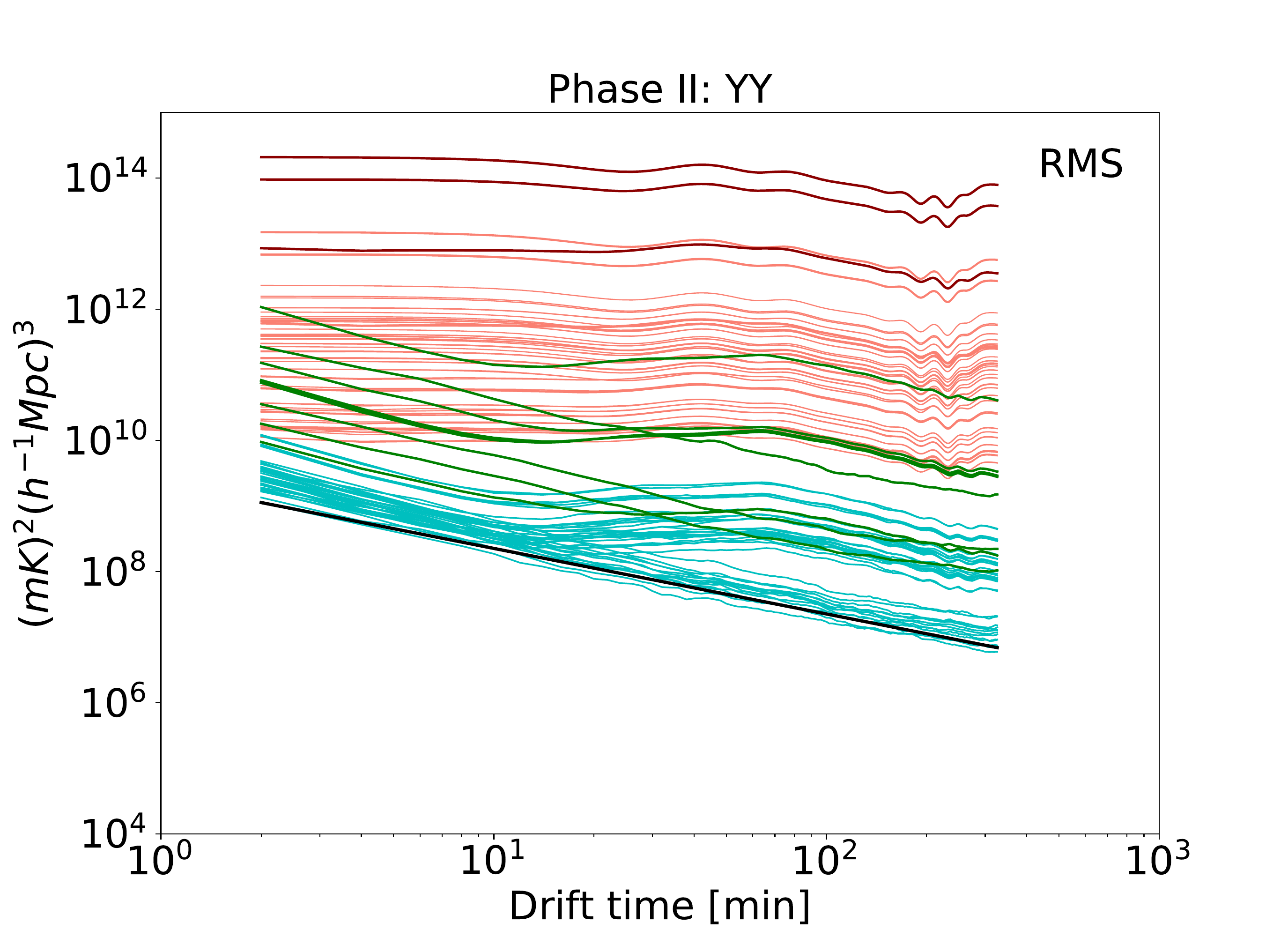}
	\caption{The panels  display the RMS of the power spectrum
          by re-defining the  power spectrum
as  the weighted average over randomly-chosen 100~voxels which yields  
54 data sets. 
The  RMS  is computed from these sets. The mean power spectrum is left unchanged by this procedure.  
The colour scheme is the same as in Figure~\ref{fig:pws_rms_plots}. The expected decrement in RMS 
for thermal noise  is approximately a  factor of 10, which is seen in the clean modes; the wedge and 
the horizontal bands which are foreground-dominated do not show this behaviour.  }
\label{fig:pws_rms_plots_100g}
\end{figure*}

In the foregoing, we performed all the  visibility cross-correlations (with the same weight)
inside a voxel and  computed the average and the RMS of this quantity over all the voxels 
(5427 voxels for a given delay parameter $\tau$).  This yields the
mean power spectrum and the RMS of  the power spectrum of a voxel as a function of drift time.  
For further noise testing,  all the  voxels  are divided into  sets of randomly-selected  100 voxels. 
This yields 54 
sets for a given $\tau$.  The power spectrum is computed for each set and then the mean 
and RMS is computed by performing weighted averaging over all the sets.  For thermal  noise, we expect 
the   RMS to  reduce by a factor of $\sqrt{100}$.   Figure~\ref{fig:pws_rms_plots_100g}  shows the RMS 
after this procedure. The ratio of the new to the old lower envelope  is
nearly a factor of 10, in consonance with   the expected decrement. We do not show the mean of the 
signal in  Figures~\ref{fig:pws_rms_plots_100g} because  the computation of the mean involves 
a linear process so it does not matter whether 
it  is computed using all the grids or first computed  over 100 grids and then
averaged over the remaining grids.

The expected decrement in the RMS after the second level of averaging  
(Figure~\ref{fig:pws_rms_plots_100g}) provides further evidence 
that a fraction of the data is uncorrupted by
either systematic errors or foregrounds and therefore is useful for the
detection of the HI signal.

Another interesting  feature of the figures  is  gradual decrease in the
power   for  the modes that are contaminated by foregrounds.  
This will be discussed in detail   in the next section.

\subsection{HI power spectrum} \label{sec:h1_powspec}
As explained above, the HI power spectrum can be estimated from the gridded
visibility data using Eq.~(\ref{eq:h1estim}) and the relations given
in the Appendix. The main input into this estimation is the function
$g(t'-t)$ which determines the time dependence of the coherence of the HI signal, as a function of baseline, 
for the primary beam of MWA (for further details see PS19). In this section
we present results for the HI power spectrum and its RMS for the combined
10-night data.

\begin{figure*}
		\includegraphics[width=0.48\textwidth]{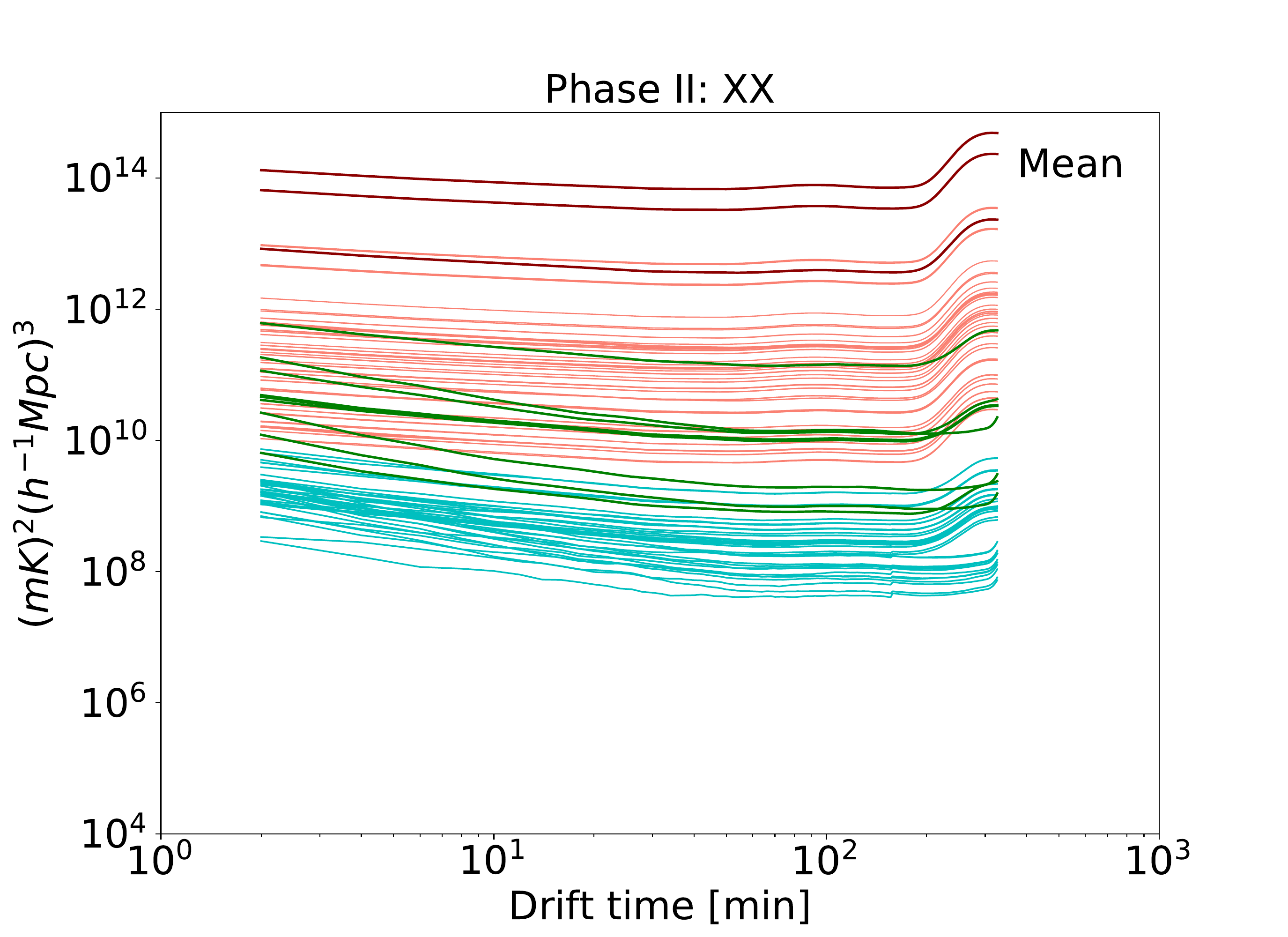}
		\includegraphics[width=0.48\textwidth]{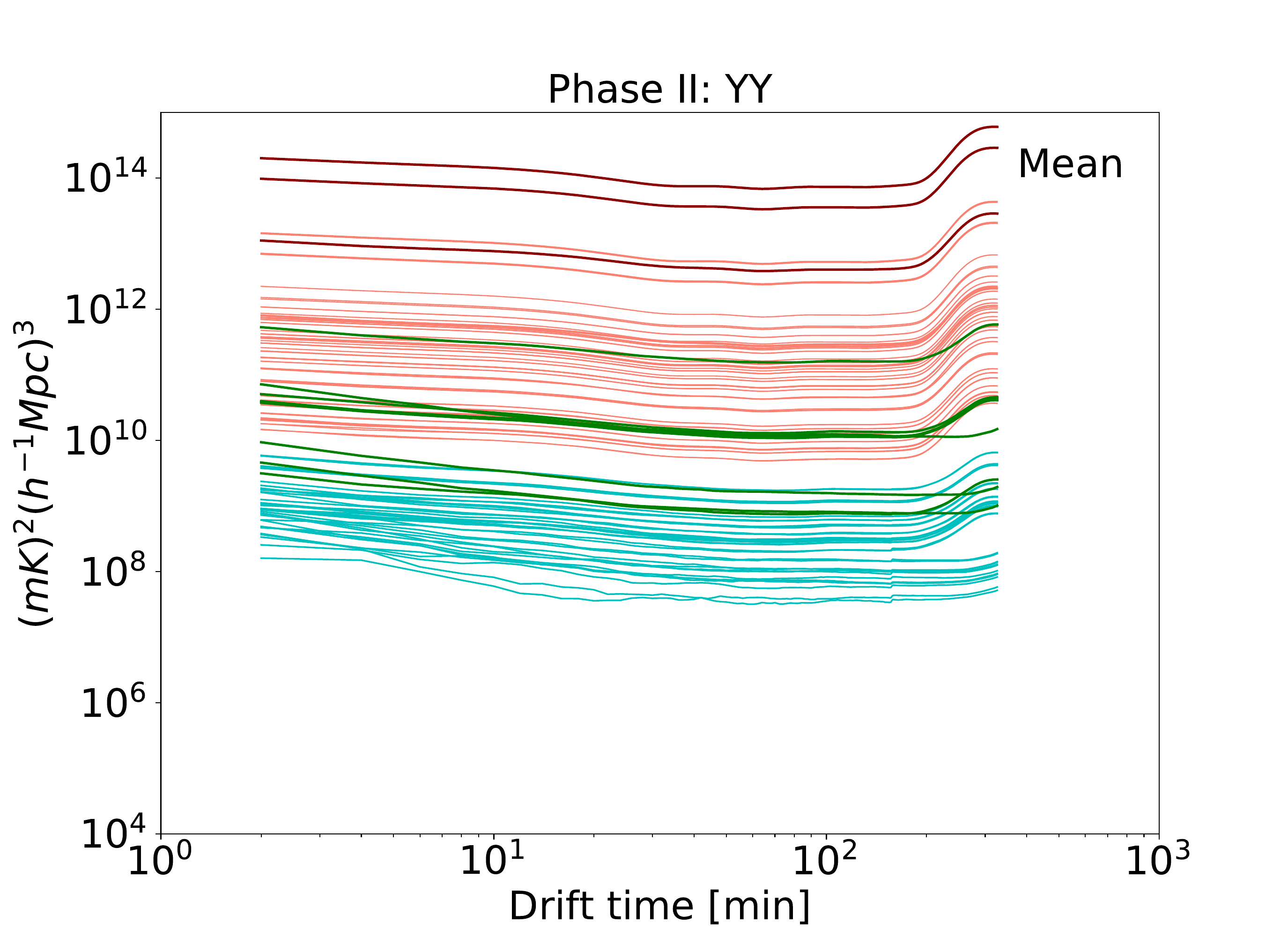}
	\caption{The mean    HI power spectra    are shown for the MWA  data as a 
	function of the drift time.   The  power spectra  are computed using  5427 voxels for a given delay parameter $\tau$. The  colour scheme is the  same as in Figure~\ref{fig:pws_rms_plots}.}
	\label{fig:pws_rms_plots_HI}
\end{figure*}
\begin{figure*}
		\includegraphics[width=0.48\textwidth]{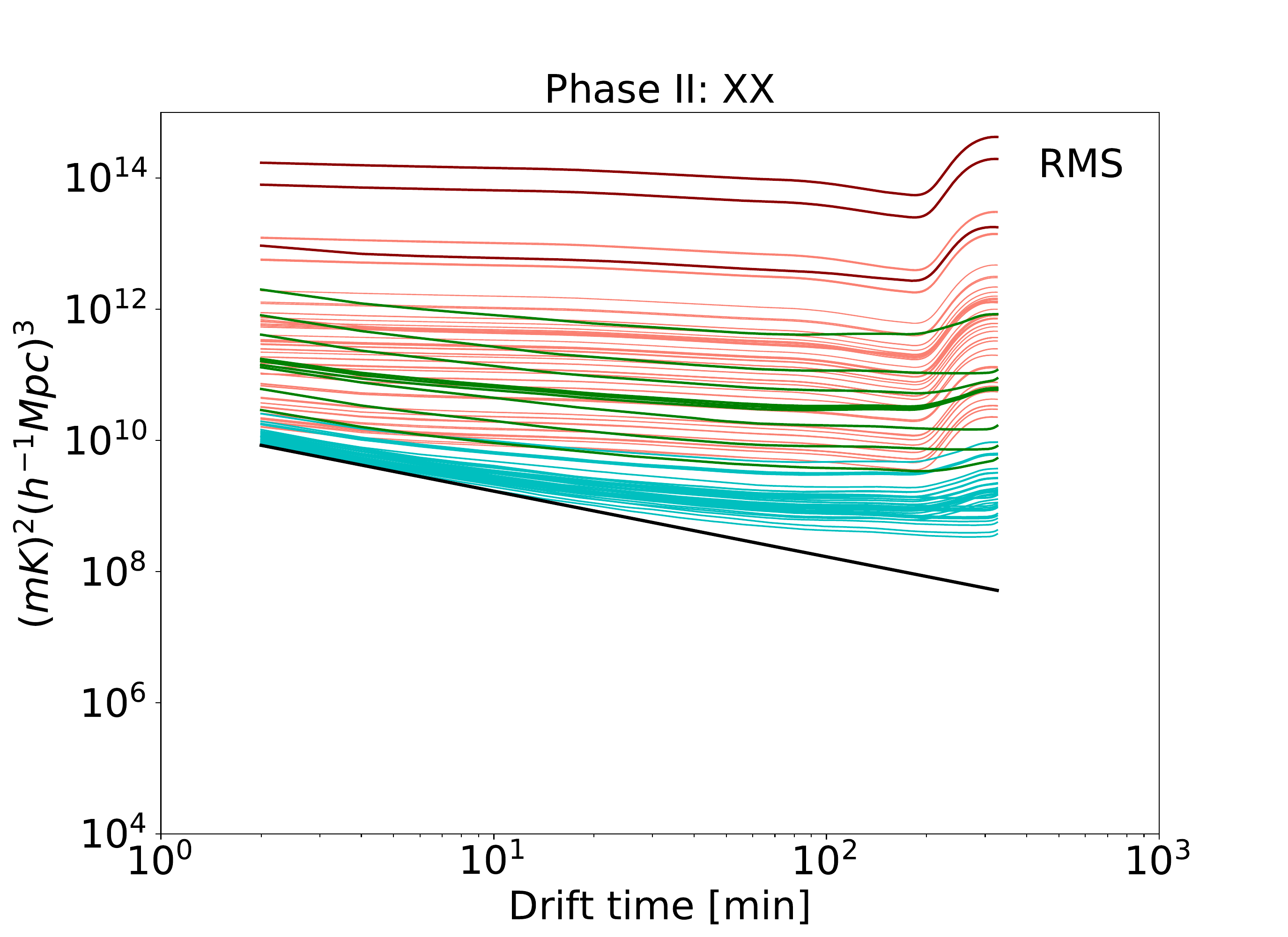}
		\includegraphics[width=0.48\textwidth]{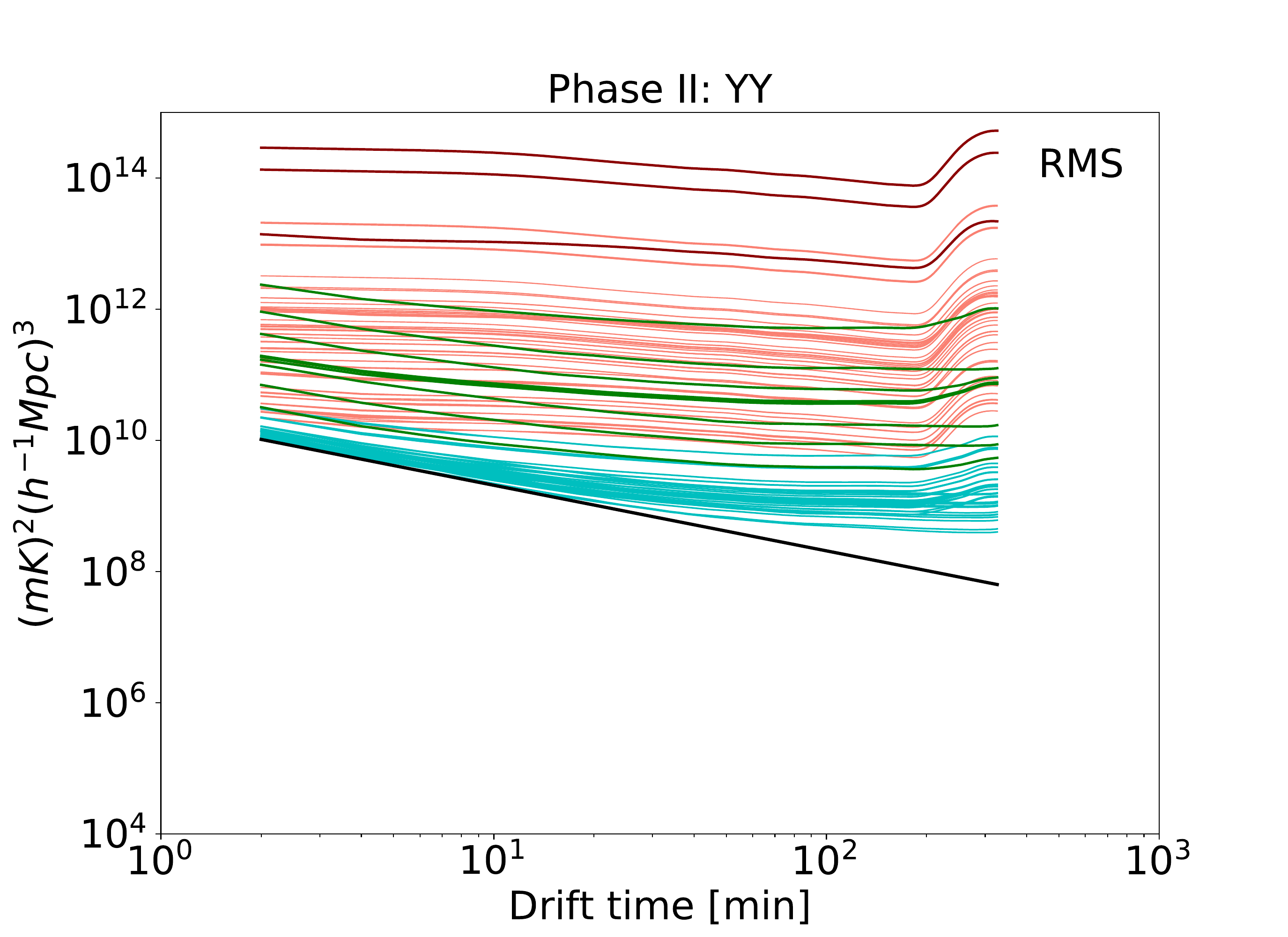}\\
		\includegraphics[width=0.48\textwidth]{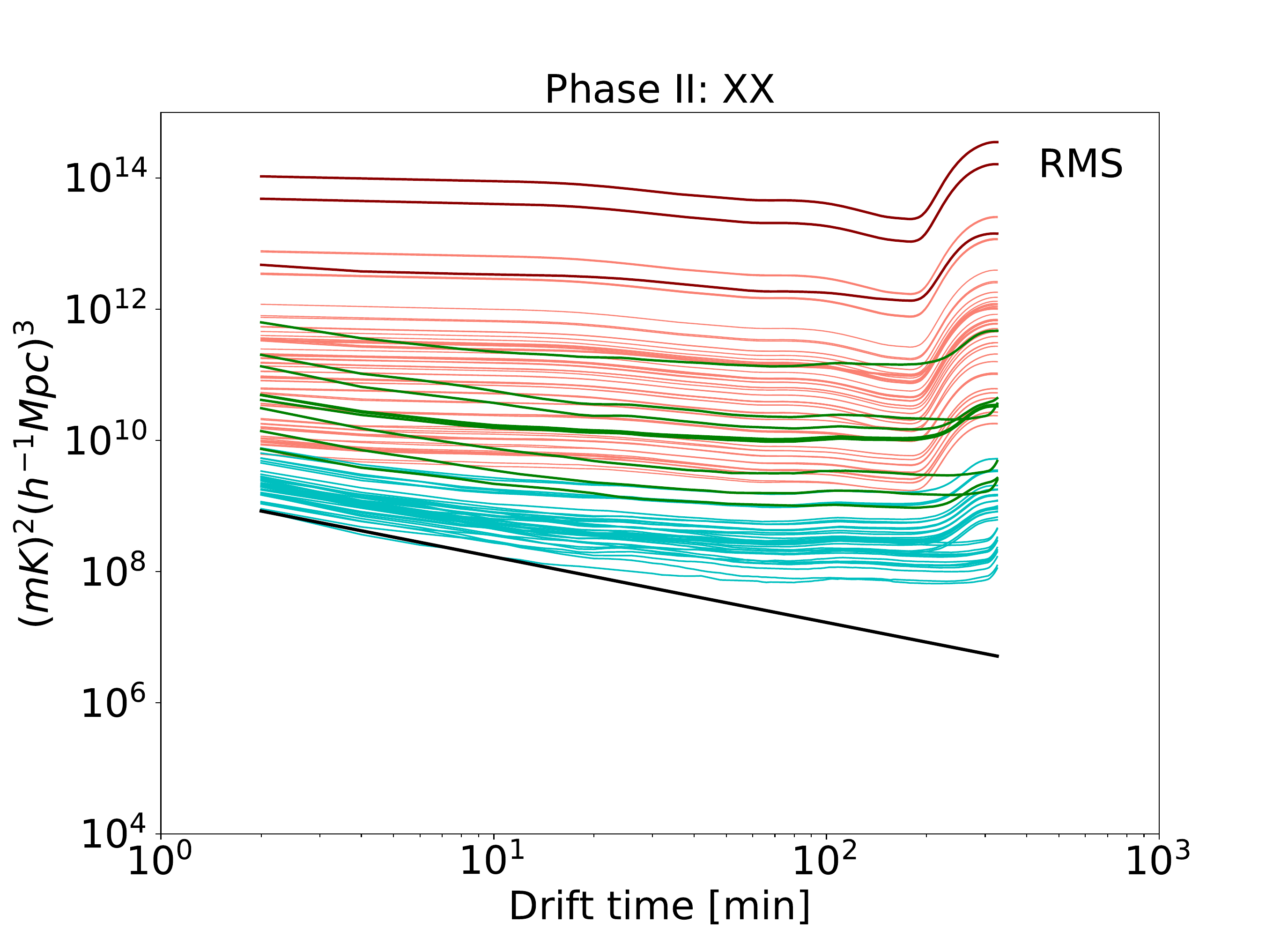}
        \centering
		\includegraphics[width=0.48\textwidth]{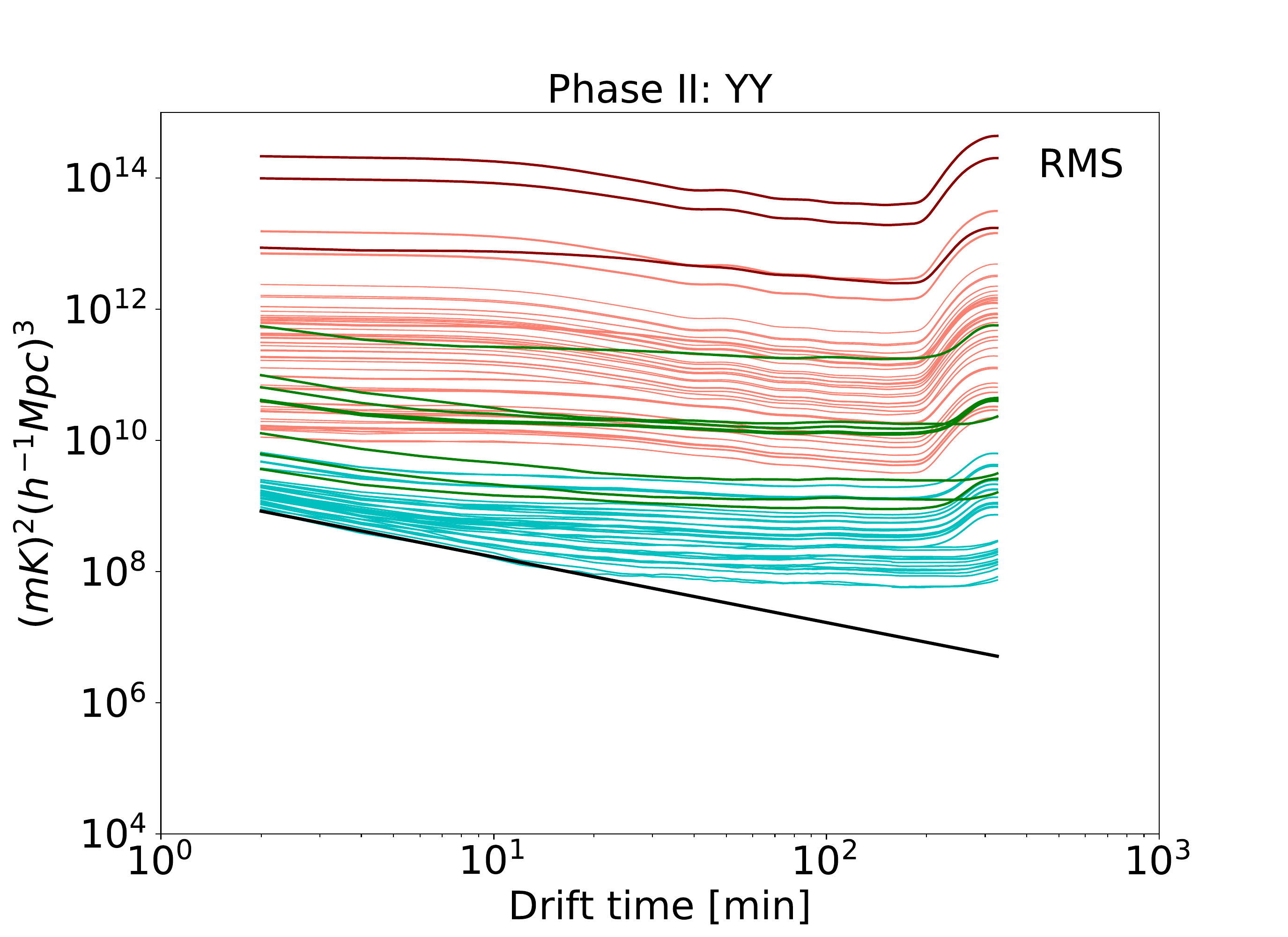}
     \caption{The top two panels show the RMS of the HI power spectrum of a voxel. 
     The RMS is computed using 5427
     voxels for a fixed $\tau$. The bottom
     two panels show the RMS after   averaging the data   over 100 randomly-chosen 
     voxels which yields  54 
     sets which are used to compute the RMS.  The mean power spectrum 
     displayed in Figure~\ref{fig:pws_rms_plots_HI} is left unchanged by this procedure. 
     The colour scheme is the same as in Figure~\ref{fig:pws_rms_plots}.  }
	\label{fig:pws_rms_plots_HI100}
\end{figure*}

In Figure~\ref{fig:pws_rms_plots_HI}, we show the mean of the HI power spectrum 
as a function of the drift time. The mean HI power spectrum
is computed by carrying out a weighted average over 5427~voxels  for a fixed
delay parameter $\tau$. 

Figure~\ref{fig:pws_rms_plots_HI100} shows the   RMS for a voxel  and
for a set of  100~voxels. The RMS of  the  power spectrum in the latter case is computed by first defining 
the power spectrum as  weighted average over 100 randomly chosen voxels (for a fixed $\tau$)  and then 
using these sets  
(54 sets) 
to compute the RMS. As noted above   the mean is left unchanged by the second 
level of averaging as it only involves linear operations on the data.    The lower envelope of  the 
Figure~\ref{fig:pws_rms_plots_100g} is plotted in the RMS plot for comparison.

Based on the discussion in the previous section, in which $g(t'-t)$ was assumed
to be unity, we can anticipate the behaviour of the the RMS of the  HI power spectrum
as a function of the drift time. Even for the shortest baselines we consider, the function  $g(t'-t)$ 
falls sharply after $t'-t \simeq 30 \, \rm min$. Therefore,  we expect the following time 
dependence of the RMS:
for a period of time for which $g(t'-t) \simeq 1$, all the visibilities inside a pixel can be
considered coherent. During this period, the RMS falls as $1/t$,  for the reasons discussed in the 
previous section. The other limiting case occurs for $t'-t$ such that $g(t'-t) \simeq 0$.
A pair of visibilities that satisfy this condition  are incoherent. In this case,
the RMS  is expected to fall as $1/\sqrt{t}$. As the period of
the drift scan far exceeds the coherence time scale of visibilities, the
time dependence of visibilities is expected to make a transition from 
$1/t$ (all visibilities coherent inside a voxel) to $1/\sqrt{t}$ (all visibilities incoherent inside a voxel). 
Figure~\ref{fig:pws_rms_plots_HI100} shows that  the   departure  of the RMS  from $1/t$ time-dependence occurs after a drift time of  nearly 20  minutes. As noted above,
for a given delay parameter, the data from all the baselines is combined (using a weighted average which depends on the density of baselines in the UV plane)  to yield the curves in Figure~\ref{fig:pws_rms_plots_HI100}. The coherence time of the HI signal varies
from a few minutes to  nearly 30~minutes for the baselines used in our analysis. For the
baseline distribution shown in  Figure~\ref{fig:ph2_uvfield}, the occupancy of a UV-pixel  is skewed in the favour of shorter baselines, which  explains  the  duration of the time of  transition seen in  Figure~\ref{fig:pws_rms_plots_HI100}. 

Finally,  a comparison between the upper and the lower panels of 
Figure~\ref{fig:pws_rms_plots_HI100} shows that the decrement in the RMS from a single voxel to 100 randomly chosen voxels 
is nearly a factor of 10 for the cleanest delay parameters even though no
such decrement is seen for the foreground-dominated modes. 
This result  provides  further proof that the clean  modes 
are noise-dominated.

\subsection{Two-dimensional power spectra and foreground wedge}
\begin{figure*}
		\includegraphics[width=0.48\textwidth]{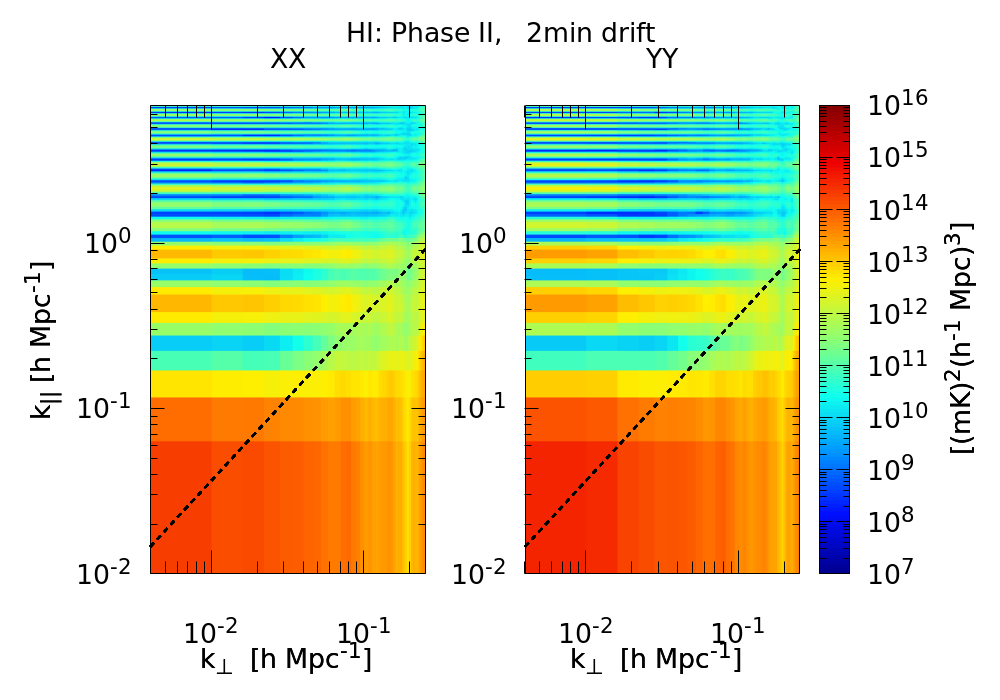}
		\includegraphics[width=0.48\textwidth]{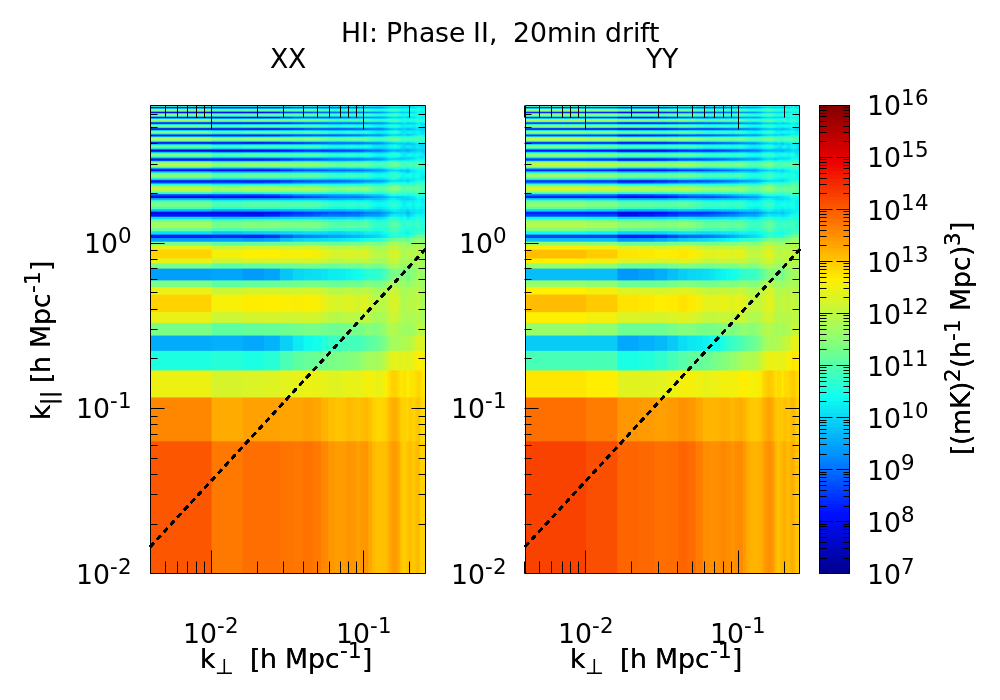}
        \caption{Two-dimensional power spectra are displayed for two fixed periods (2 and 20 minutes) 
          of drift scans. The diagonal black line (plotted using  the main lobe of MWA's primary beam) separate the EoR window (the region above the line) from the foreground-dominated modes.}
  \label{fig:2d_120min_ph2}
\end{figure*}

\begin{figure*}
		\includegraphics[width=0.48\textwidth]{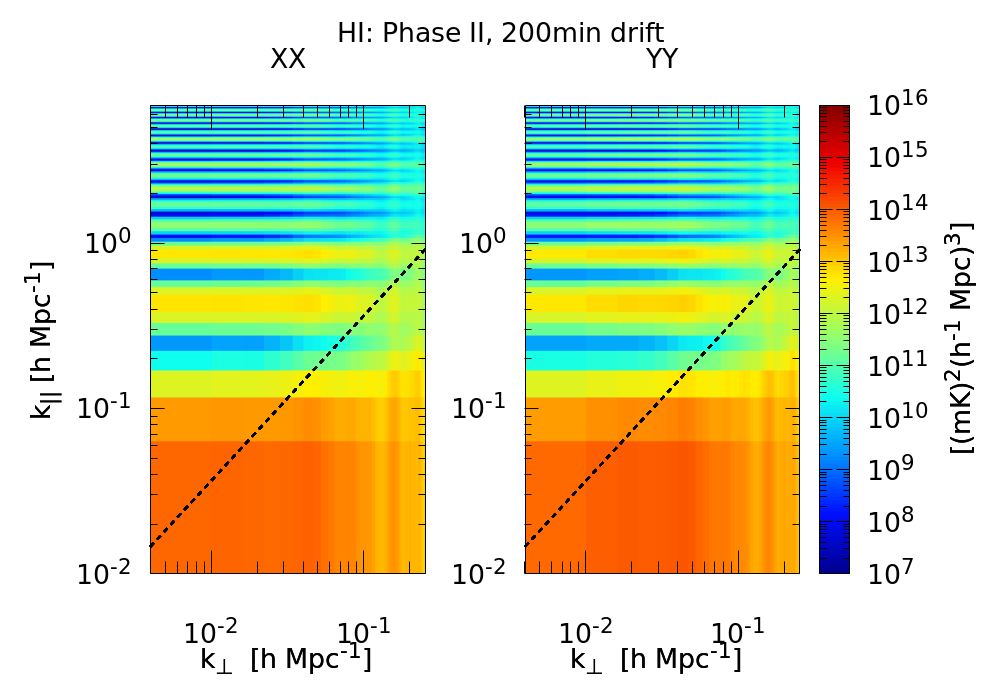}
		\includegraphics[width=0.48\textwidth]{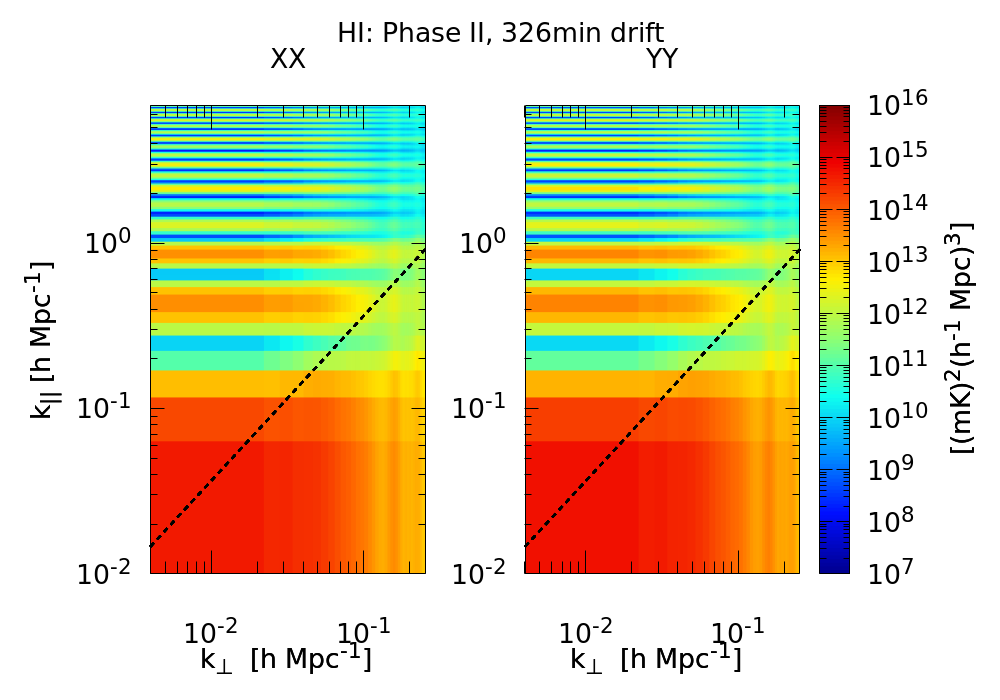}
\caption{The 10-night combined data are displayed for 200~minutes and 326~minutes of drift scan.}
  \label{fig:2d_200min_ph2}
\end{figure*}

Figures~\ref{fig:pws_rms_plots_HI} and~\ref{fig:pws_rms_plots_HI100} adequately capture the time-dependence
of the  power spectra for different delay parameters  and the separation of foreground-dominated
modes from  noise-dominated modes. To further analyze the complex structure of the
signal in the Fourier domain, we show the signal in the usual   $k_{\perp}\hbox{--}k_{\parallel}$ domain in 
Figures~\ref{fig:2d_120min_ph2} and~\ref{fig:2d_200min_ph2}.  In these figures, cylindrical averaging 
for a  fixed $k_\perp = |{\bf k_\perp}| = \sqrt{ k^{2}_{\perp1} + k^{2}_{\perp2}}$ is performed.


We notice  the characteristic features of delay space power spectra for MWA:  foreground dominated wedge, 
cleaner EoR window, and the   horizontal bands owing to the missing  MWA spectral
channels. 
As we already noted above, to reduce power leakage between  $k_{\parallel}$ bins  we apply 
Blackman-Nuttall window on visibilities before taking Fourier transform along the frequency axis 
(Eq.~(\ref{BN_win})). While the application of this  window reduces the leakage
 and therefore the  EoR 
window is cleaner, it  thickens the horizontal bands. 

The two-dimensional power spectrum should be viewed alongside  Figure~\ref{fig:pws_rms_plots_HI}. 
Figure~\ref{fig:pws_rms_plots_HI} shows the mean power spectrum computed by averaging over all the 
baselines of fixed $k_\parallel$ while the two-dimensional power spectra provide  additional  information  
for a given   $|{\bf k_\perp}|$. The time dependence in both the cases is similar:
as the drift scan time is increased from 2~minutes to 
120~minutes, the  EoR window gets cleaner by up to  two orders of magnitude owing to the reduction of 
noise.

\subsection{Foregrounds in drift scan}
The separation of foreground-dominated modes from noise-dominated modes is
a key requirement of the  EoR science. 
The focus of this subsection is a discussion of the  modes dominated by foregrounds with particular focus on drift scans but a part of the discussion is general enough to be applicable to the  tracking observation also.

As shown in PS19, the behaviour of foregrounds in a drift scan could be
markedly different from tracking observations (for another perspective on foregrounds in a drift 
scan see \cite{m-mode1}). The coherence time scale
of the HI signal is larger than that of the extragalactic point
sources  while  it is comparable to the coherence time scale of the
diffuse sources (modelled as a statistically homogeneous  process in PS19). This
means point sources can get uncorrelated in the process of extracting the
HI signal.

We first consider  Figure~\ref{fig:pws_rms_plots}  which  assumes  $g(t'-t) = 1$. In this case, 
all the cross-correlations within a voxel are assigned  equal weight. As the time of the scan is much 
larger than the coherence time scale of all the components, we expect partial de-correlation of point sources, 
the HI signal, and
diffuse foregrounds. We notice a decline of power by  nearly an order of magnitude
in the foreground-dominated modes until late in the scan when Fornax~A and the galactic plane start 
contributing significantly. This behaviour should be compared
to that in Figure~\ref{fig:pws_rms_plots_HI} in which $g(t'-t)$ corresponds to  the
coherence function for HI signal for MWA. In this case, only a fraction of
cross-correlations inside a voxel are carried out, which prevents the de-correlation of the HI signal. 
In this case, the decrement of the power in foreground
dominated modes is also shallower because this process  also prevents the de-correlation of 
diffuse foregrounds.

The complexity of the time-dependence of  foregrounds in a drift scan is further
revealed in Figures~\ref{fig:2d_120min_ph2} and~\ref{fig:2d_200min_ph2}. In these figures, the  power  
in Fourier modes in the plane of the sky is separated from the modes
along the line of sight. Unlike Figure~\ref{fig:pws_rms_plots_HI} which displays the  power spectra 
averaged over all $|{\bf k_\perp}|$ for a given $k_{\parallel}$, 
Figures~\ref{fig:2d_120min_ph2} and~\ref{fig:2d_200min_ph2} show the results for 
a fixed $|{\bf k_\perp}|$. This allows us to
discern the baseline dependence of the de-correlation process in a drift scan.
From Figure~1 of PS19, we note that the de-correlation time of the HI signal
varies from nearly 30~minutes to 5~minutes  from the shortest to the longest
baselines we consider in our study, $\sqrt{u^2+v^2} \simeq 20\hbox{--}300$.

Figure~\ref{fig:2d_120min_ph2}  clearly show the
depletion of power in foreground-dominated modes  as the drift time increases from 2~minutes to 
20~minutes.  However, this process does not cause the transition of
any foreground-dominated
mode to a noise dominated mode. This allows us to conclude that, for  the data we analyse, drift scans don't  add any further information on the separation
of foregrounds from the EoR window as compared to the tracking observation.

To get further evidence of the noise-dominated modes, we focus on Figure~\ref{fig:pws_rms_plots_HI100}. This figure shows the RMS of the HI power spectrum  decreases  by the expected factor
of nearly ten after the second level of averaging for the noise-dominated modes. However, this decrement is not seen  for the foreground-dominated modes. Figures~\ref{fig:pws_rms_plots} and~\ref{fig:pws_rms_plots_100g} also support this conclusion. This constitutes another diagnostic  of the noise-dominated modes which are suitable for the EoR detection. We note that this method of comparing the  RMS
of the HI power spectrum using different sub-sets of data provides   a suitable diagnostic  for  both  the  drift scan and tracking observations.

\subsection{One-dimensional power spectrum}
The information displayed in Figures~\ref{fig:pws_rms_plots_HI} and~\ref{fig:2d_200min_ph2} 
can be partially  summarized with  one-dimensional power spectrum defined as: 
$\Delta^2(k) = k^3P(k)/(2\pi^2)$ with $k=\sqrt{{\bf k}_\perp^2 +k_\parallel^2}$.  
For the EoR window, MWA baseline distribution gives  $k_\parallel \gg |{\bf k}_\perp|$ and therefore 
the one-dimensional power spectrum can  be computed by  averaging over all ${\bf k}_\perp$ 
for a fixed $k_\parallel$ (it is partly the motivation of the choice of variables for 
Figure~\ref{fig:pws_rms_plots_HI}).  
The one-dimensional power spectrum  as a function of time 
can be computed  from  Figure~\ref{fig:pws_rms_plots_HI} (or Figure~\ref{fig:2d_200min_ph2}). The best results
is obtained after  nearly 200~minutes of the scan\footnote{To avoid
  confusion it should be underlined that the total amount of data for
  200~minutes of scan is the combined data from ten nights and therefore
  correspond to nearly 2000~minutes of data from multiple scans over the same field.},
as clearly seen in
Figure~\ref{fig:2d_200min_ph2}. The  contamination from  
Fornax~A and the galactic plane prevents any further improvement. This also means that only 35~hours of data
from a total of 55~hours is usable for our purposes. 

Figure~\ref{fig:2d_200min_ph2} shows that the largest scales in the EoR window  that 
can be probed with MWA correspond to  $k \simeq 0.2$\kunits, which gives us  
$\Delta^2(k) \simeq (1000~\rm mK)^2$ (for either XX or YY polarization). 
Figure~\ref{fig:2d_200min_ph2} also shows that the cleanest modes are obtained for much smaller
scales. For $k\simeq 1$\kunits,  the one-dimensional power spectrum  
$\Delta^2(k) \simeq (1000~\rm mK)^2$,  which is comparable to the value at larger scales.

The RMS of the HI power spectrum can be computed from Figure~\ref{fig:pws_rms_plots_HI100}. 
It is based on 54 
sets with each set obtained from  weighted average over 100~voxels. 
This yields an RMS of $\simeq  10^6 \, \rm (mK)^2$ for
clean modes in the range $k \simeq 0.02 \hbox{--}1$\kunits.
The RMS estimate can be further improved by averaging over all the  data
for a fixed $\tau$ (this leaves the mean unchanged for reasons outlined in the foregoing) and then
computing the RMS using bootstrapping (e.g. \cite{paper_reanalysis}).

We plot  the power spectra for both the 
XX and YY polarizations in all the  figures 
to emphasize the long-term  system stability in a drift scan. 
Our calibration does not involve a polarized source so both the polarizations 
are assigned equal weight in the beginning of the scan. Our results show  that 
no significant deviation emerges at the end of the scan. 
As the HI signal is unpolarized, the power spectra for the two polarizations
can be added in quadrature to yield a further improvement in RMS by  nearly a factor  of $\sqrt{2}$.

\subsubsection{Comparison with other HI  power spectrum measurements}
Our results can be directly compared to PAPER  HI power spectra upper limits 
as it was also  a drift scan experiment. 
{PAPER was located in Karoo desert, South Africa at coordinates 
$30^\circ43'17''$S, $21^\circ25'42''$E.
Its 2-meter antenna had a primary beam (FWHM) of nearly $50^\circ$. PAPER's 64 antennas were placed
in a 8x8 pattern. For the final results presented in \cite{paper_reanalysis},  6 of the antennas were
flagged and only a subset of baselines were used for  the power spectrum estimations. 
This subset consisted of 30-meter East-West baselines 
between adjacent columns  and 
those staggered by one row (see Figure~2 of \citealt{paper_reanalysis}).
The system temperature for the observational runs varied between 400--800~K in the frequency range.
These data sets had a total bandwidth of 10~MHz with a channel
width of 493~kHz, centered around six frequencies in the range 120--167~MHz. 
The power spectrum estimate was based on 135~days of observations, but only the data from 124~days was
used for  LST binning. The LST spanned the range  
$00^{\rm h}30^{\rm m}00^{\rm s}$ to $08^{\rm h}36^{\rm m}00^{\rm s}$. The data sets were initially binned
with a time resolution 43~seconds. After the LST binning, they were 
further averaged with a top-hat filter in time
which yielded   a resolution of  938~seconds. 
With these instrumental parameters, the PAPER collaboration recently published the 
analysis of its final data products.
Their upper limits on the HI power spectra at $z\simeq 8.13$ are:
$\Delta^2(k) \simeq(1000~{\rm mK})^2$ at $k = 0.2$\kunits and 
$(380~{\rm mK})^2$ at $k \simeq 0.32$\kunits (\citealt{paper_reanalysis}). 
Our upper limit at  a  comparable wavenumber and redshift  is  $\simeq(1000~{\rm mK})^2$,  
based on nearly 35~hours of MWA drift scan data.

While a  more direct  comparison between the performance of MWA and PAPER is harder, 
this allows us to  note the following
salient differences between the two interferometers and  analysis pipelines: (a) the collecting area of MWA is nearly five times large, (b) PAPER baselines have a higher level of redundancy  (section~\ref{sec:data_meth}), (c) the time over which the visibilities are averaged is around 15~minutes for PAPER while it varies between a few minutes to nearly 30~minutes for MWA (for details of coherence time of these
interferometers see PS19),  (d) the drift  time for PAPER is nearly 40 times larger as compared to
our run, (e)  cross-channel contamination is an important  issue for MWA owing to missing channels which prevents a large fraction of  modes from reaching the theoretical noise levels (Figure~\ref{fig:pws_rms_plots}). }

More recently, MWA collaboration published results from the analysis of  its best tracking Phase~I and
Phase~II data 
for three fields, EoR0, EoR1, and EoR2 for a range of redshifts and
wavenumbers  \citep{2020MNRAS.493.4711T}. 
{The integration time on fields relevant to us
varies from 27 to 38~hours which is comparable to the
drift time in this paper. }
Even though a direct comparison is difficult
owing to different modes of observations, our analysis is based on data
that traverses between EoR0 and EoR1 fields, and therefore a comparison of our
results with tracking data from EoR0 and EoR1 fields for the same   
redshift  range and wavenumbers is justified.

For $z\simeq 8.2$ and $k \simeq  0.2$\kunits, the following upper limits were
obtained by the MWA tracking observation: $(376~{\rm mK})^2$ (EoR0, 38~hours)
and $(1166~{\rm mK})^2$ (EoR1, 27~hours). The limit we obtain in this paper
is worse than the limit for the  EoR0 field but marginally better than the
limit for the EoR1 field.  
{ While the analysis of the tracking data reports the power spectrum  for
$k < 0.7$\kunits, the primary results for the tracking data are more reliable  for  
$k < 0.4$\kunits \citep{2020MNRAS.493.4711T}. As noted above, our best results are obtained for the 
scale $k \simeq 1$\kunits. The analysis of the tracking data yielded the }
following upper limit 
for comparable scales: $\Delta^2(k) \simeq(2000~{\rm mK})^2$ (EoR0) and
$\Delta^2(k) \simeq(3100~{\rm mK})^2$ (EoR1). Our results are an improvement
over the tracking observation at this scale.

\section{Summary and conclusion} \label{sec:concl}
The use of drift scan data to extract the HI power spectrum from high and
low redshift data is an established method
(e.g. \cite{chime,paper_reanalysis,fringe-rate,deboer17}).  Drift scans are expected to yield   
superior system stability which  is one of the key requirements for the detection of the weak HI  signal.
In this paper we report  the analysis of nearly 55 hours of   publicly-available MWA Phase~II drift 
scan  EoR  data. Our analysis is based on a  novel method proposed
in PS19, which is an extension of formalism given by \cite{paul14}. We develop
a pipeline which works in two modes: (a) noise testing:  the aim of this mode is to
test system stability by  comparing  the data power spectrum against 
uncorrelated noise as a function of the drift time, (b) HI mode: the HI power spectrum is computed 
in this mode. 
We summarize our main results and findings:

\begin{itemize}
\item {\it Noise testing}: Figures~\ref{fig:pws_rms_plots} and~\ref{fig:pws_rms_plots_100g} 
show the main results. The figures demonstrate that the data
  agree with the behaviour of the  thermal  noise for  a drift scan of
  nearly  4~hrs---the RMS falls as $1/t$ during this period.  This  provides reasonable  proof that 
  the system parameters (primary beam, bandpass) are stable over the duration of the scan. This test 
  also allows us to estimate the mean system temperature during the scan, which is in agreement 
  with the reported values.

\item {\it HI power spectrum}: The HI power spectra as a function of time  are shown in
  Figures~\ref{fig:pws_rms_plots_HI} and~\ref{fig:pws_rms_plots_HI100}.
  The two-dimensional plots (Figures~\ref{fig:2d_120min_ph2} and~\ref{fig:2d_200min_ph2}) show the HI 
  power spectrum in $k_{\perp}\hbox{--}k_\parallel$ plane for fixed drift  times. These results are in line
  with the expectation that, for the clean modes,  the RMS of the mean HI power spectrum initially falls as $1/t$ and then approximately as 
  $1/\sqrt{t}$. This transition  occurs when the drift time exceeds the coherence time of the HI signal. A comparison between the RMS
  computed  from two different data subsets
  (Figure~\ref{fig:pws_rms_plots_HI100}) also shows  that the clean
  modes are noise-dominated and therefore suitable for EoR studies. 
\item{\it Foreground-dominated modes}: As shown in PS19, the point sources
  de-correlate on time scales much shorter than the HI signal, which
  means we expect some level of decrement in the power of the  foreground-dominated modes even as 
  the HI signal is extracted from the data. This is seen
  in Figures~\ref{fig:pws_rms_plots_HI} and~\ref{fig:pws_rms_plots_HI100}. 
  However, it is difficult to draw a definite conclusion regarding the depletion of
  the power owing to the complicated nature of diffuse foregrounds.
\end{itemize}

The aim of this paper is to demonstrate   a new method of analysing  MWA drift 
scan  EoR data. This method is particularly suited for
repeated scans over a given field. We seek a proof of concept  by testing system stability during a 
scan. We also compute the HI power spectrum. 

The detection of the HI signal from the epoch of reionization remains a challenge. 
Given the weak   HI signal buried under  strong  foregrounds and  hundreds of hours of integration 
time needed to reduce the thermal noise to acceptable levels, it is perhaps imperative that 
multiple approaches are  employed to
understand and analyse the signal.

\section*{Acknowledgements}
This scientific work makes use of the Murchison Radio-astronomy Observatory, operated by CSIRO. 
We acknowledge the Wajarri Yamatji people as the traditional owners of the Observatory site. 
Support for the operation of the MWA is provided by the Australian Government (NCRIS), 
under a contract to Curtin University administered by Astronomy Australia Limited. 
We acknowledge the Pawsey Supercomputing Centre which is supported by 
the Western Australian and Australian Governments.

\section*{Data Availability}
The data sets were derived from sources in the public domain
(the MWA Data Archive: project ID G0031) at \url{https://asvo.mwatelescope.org/}.



\bibliographystyle{mnras}
\bibliography{bib.bib}

\appendix
\section{}\label{sec:appen}
Here we  describe briefly the conversion of measured quantity (Eq.~(\ref{eq:h1estim})) 
to the variables of the HI signal.

The parameters  of the radio interferometer can be related to the Fourier
variables  of the   HI signal as (e.g. \cite{paul16} and references therein):
\begin{equation}
k_{\perp 1} = \frac{2\pi}{r_{0}} u_{0}, k_{\perp 2} = 
\frac{2\pi}{r_{0}} v_{0}, k_{\parallel} = \frac{2\pi}{|\dot{r}_{0}|} \tau.
\end{equation}
Here $k_{\perp 1}$ and $k_{\perp 2}$ are the  Fourier components on the sky plane while 
$k_{\parallel}$ lies along the line of sight. $r_0$ is the coordinate
distance corresponding to the observed redshifted frequency  $\nu_0$ 
and  $\dot{r}_{0} = dr/d\nu$ at $\nu=\nu_0$.

The HI power spectrum can be written in terms of the visibility correlation function defined in 
Eq.~(\ref{eq:h1estim}) using:
\begin{align}\label{eq:defpwsapprox2}
{P}_{\rm HI}(k) \simeq \frac {9 |\dot{r}_{0}| r^{2}_{0}} {4 \bar{I}^{2}_{0} B \Omega_{0} } 
|{\cal C}_\tau(\textbf{u}_{0},w_0)|.
\end{align}
Here  $\Omega_{0} = \frac{\lambda^2_{0}}{ A_{\rm eff}}$ is the solid angle  of the primary beam of MWA 
(the effective area of an MWA tile, $A_{\rm eff} = 21.5 \, \rm m^2$ at $\nu = 154 \, \rm MHz$; 
for details see \citealt{tingay13_mwasystem}) and   
$ k = \sqrt{ k^2_{\perp 1} + k^2_{\perp 2} + k^2_{\parallel} }$. $\bar{I}$ is the mean intensity of 
the HI signal. For a single polarization (either XX or YY correlation), 
$\bar{I}_{0} = k_{B}T_{B}/\lambda^{2}_{0}$, where $T_B$ is the  mean HI brightness temperature and  
$k_B$ is the Boltzmann constant. For computing the mean intensity, we assume a  
neutral hydrogen fraction of  0.5 at $z = 8.21 ~(154.24\text{MHz})$, which yields  
$T_B = 12.6~\rm mK$. This gives us the normalization needed to convert visibility correlation 
from $(\rm JyHz)^2$ to \pkunits
, the units of the HI power spectrum:
\begin{equation}
\frac{9 \lambda^{4}_{0} |\dot{r}_{0}| r^{2}_{0}}{4 k^{2}_{B} B \Omega_{0} } = 4.18\times10^{-3} 
\frac{ \text{\pkunits} } {(\rm JyHz)^2}. \label{eq:h1norm} 
\end{equation}
The factor of 9/4 in the  normalization is specific to the MWA primary beam.
We use Eqs.~(\ref{eq:h1norm}) and~(\ref{eq:defpwsapprox2}) for the analysis of the  data.

\bsp	
\label{lastpage}
\end{document}